\documentclass[aps,preprint,superscriptaddress,nofootinbib,preprintnumbers,eqsecnum,prd]{revtex4-1}

\usepackage{bm,amssymb,slashed,graphicx,multirow,soul,mathtools}

\newcommand{\widebar}[1]{
	\mathchoice{}
	{	\smash{\overset{\mbox{\underline{{\phantom{n}}}}}{\smash{#1}\phantom{\rule{0.1em}{1.4ex}}}}\hspace{-0.2em}
	}{	\smash{\overset{\mbox{\underline{{\phantom{\scriptsize{n}}}}}}{\smash{#1}\phantom{\rule{0.05em}{1.05ex}}}}\hspace{-0.1em}
	}{	\smash{\overset{\mbox{\underline{{\phantom{\scriptsize{n}}}}}}{\smash{#1}\phantom{\rule{0.05em}{0.7ex}}}}\hspace{-0.1em}
	}
}
 
\newcommand{\dds}{D^{(*)}}
\newcommand{\ddsbar}{\bar{D}^{(*)}}
\newcommand{\bbar}{\bar{b}}
\newcommand{\cbar}{\bar{c}}
\newcommand{\Bbar}{\bar{B}}
\newcommand{\Dbar}{\bar{D}}
\newcommand{\Wbar}{\widebar{W}}
\newcommand{\tnbar}{{\bar{\nu}_\tau}}
\newcommand{\tn}{\nu_\tau}
\newcommand{\bdtn}{{B \to D \tau \tn}}
\newcommand{\bdstn}{{B \to D^{*}\tau \tn}}
\newcommand{\bddstn}{{B \to D^{(*)}\tau \tn}}
\newcommand{\mn}{{\mu\nu}}
\newcommand{\g}{\gamma}
\newcommand{\ampB}[2]{\big\langle #1 \big|\, #2\, \big| B \big\rangle }
\newcommand{\ampBb}[2]{\big\langle #1 \big|\, #2\, \big| \Bbar \big\rangle }
\newcommand{\ampBbs}[2]{\langle #1 |\, #2\, | \bar{B} \rangle }
\newcommand{\dotpr}[2]{\,#1\!\cdot #2\, }
\newcommand{\epspr}[4]{\epsilon^{\,#1 \,#2 \,#3 \,#4}}
\newcommand{\Pw}{|\bm{q}^*|}
\newcommand{\Pp}{|\bm{p}_{\pi}^*|}
\newcommand{\PS}{\mathcal{PS}}

\def\spnt{1}
\if\spnt1
	\newcommand{\up}{+}
	\newcommand{\dn}{-}
\else
	\newcommand{\up}{2}
	\newcommand{\dn}{1}
\fi		

\newcommand{\alSL}{\alpha_L^S}
\newcommand{\alSR}{\alpha_R^S}
\newcommand{\alVL}{\alpha_L^V}
\newcommand{\alVR}{\alpha_R^V}
\newcommand{\alTL}{\alpha_L^T}
\newcommand{\alTR}{\alpha_R^T}
\newcommand{\beSL}{\beta_L^S}
\newcommand{\beSR}{\beta_R^S}
\newcommand{\beVL}{\beta_L^V}
\newcommand{\beVR}{\beta_R^V}
\newcommand{\beTL}{\beta_L^T}
\newcommand{\beTR}{\beta_R^T}

\newcommand{\thtau}{\theta_{\tau}}
\newcommand{\phtau}{\phi_{\tau}}
\newcommand{\thW}{\theta_{W}}
\newcommand{\phW}{\phi_{W}}
\newcommand{\thell}{\theta_{\ell}}
\newcommand{\phell}{\phi_{\ell}}
\newcommand{\thpi}{\theta_{\pi}}
\newcommand{\phpi}{\phi_{\pi}}
\newcommand{\thD}{\theta_{D}}
\newcommand{\phD}{\phi_{D}}
\newcommand{\phDtau}{(\phD - \phtau)}

\newcommand{\DpiA}{\Delta}
\newcommand{\DpiB}{\Sigma}
\newcommand{\DgaA}{\Omega}
\newcommand{\DgaB}{\Xi}

\def\lqcd{\Lambda_\text{QCD}}

\makeatletter
\g@addto@macro\bfseries{\boldmath}
\makeatother

\allowdisplaybreaks
\graphicspath{{./Figures/}}

\usepackage{hyperref}

\begin{document}

\title{New Physics in the Visible Final States of \texorpdfstring{$B\to D^{(*)}\tau\nu$}{bdtn}}

\author{Zoltan Ligeti}
\author{Michele Papucci}
\affiliation{Ernest Orlando Lawrence Berkeley National Laboratory, University of California, Berkeley, CA 94720, USA}
\affiliation{Berkeley Center for Theoretical Physics, Department of Physics, University of California, Berkeley, CA 94720, USA}

\author{Dean J.\ Robinson}
\affiliation{Ernest Orlando Lawrence Berkeley National Laboratory, University of California, Berkeley, CA 94720, USA}
\affiliation{Berkeley Center for Theoretical Physics, Department of Physics, University of California, Berkeley, CA 94720, USA}
\affiliation{Physics Department, University of Cincinnati, Cincinnati OH 45221, USA}

\begin{abstract}
We derive compact expressions for the helicity amplitudes of the many-body $B \to D^{(*)}(\to DY)\tau(\to X\nu)\nu$ decays, specifically for $X = \ell \nu$ or $\pi$ and $Y = \pi$ or $\gamma$.  We include contributions from all ten possible new physics four-Fermi operators with arbitrary couplings.  Our results capture interference effects in the full phase space of the visible $\tau$ and $D^*$ decay products which are missed in analyses that treat the $\tau$ or $D^*$ or both as stable. The $\tau$ interference effects are sizable, formally of order $m_\tau/m_B$ for the standard model, and may be of order unity in the presence of new physics. Treating interference correctly is essential when considering kinematic distributions of the $\tau$ or $D^*$ decay products, and when including experimentally unavoidable phase space cuts. Our amplitude-level results also allow for efficient exploration of new physics effects in the fully differential phase space, by enabling experiments to perform such studies on fully simulated Monte Carlo datasets via efficient event reweighing. As an example, we explore a class of new physics interactions that can fit the observed $R(D^{(*)})$ ratios, and show that analyses including more differential kinematic information can provide greater discriminating power for new physics, than single kinematic variables alone.
\end{abstract}

\maketitle
\tableofcontents
\clearpage

\section{Introduction}
Over the past few years, the BaBar~\cite{Lees:2012xj, Lees:2013uzd}, Belle~\cite{Huschle:2015rga, Abdesselam:2016cgx, Abdesselam:2016xqt} and LHCb~\cite{Aaij:2015yra} experiments have reported a persistent anomaly in the ratios
\begin{equation}
	R(\dds) \equiv \frac{\Gamma[\bddstn]}{\Gamma[B \to \dds \ell \nu]}\,,\qquad \ell = \mu,\,e\,,
\end{equation}
compared to the standard model (SM) expectations.  The latter are fairly precise, because heavy quark symmetry~\cite{Isgur:1989vq, Isgur:1989ed, Manohar:2000dt} and data constrain the $B \to \dds$ form factors.  The world averages for $R(\dds)$~\cite{Amhis:2014hma} show a tension with the SM at approximately the $4\sigma$ level, motivating consideration of possible new physics (NP) contributions to this signal.

Signatures of NP in $B \to X \tau \tn$ are of long-standing interest (see e.g. Refs~\cite{Krawczyk:1987zj,Kalinowski:1990ba,Hou:1992sy,Goldberger:1999yh,Nierste:2008qe}), and a large number of recent studies~\cite{Lees:2012xj, Lees:2013uzd,Huschle:2015rga, Abdesselam:2016cgx, Abdesselam:2016xqt,Aaij:2015yra,Fajfer:2012vx,  Sakaki:2012ft, Fajfer:2012jt,Crivellin:2012ye,Datta:2012qk,Celis:2012dk, Tanaka:2012nw, Biancofiore:2013ki,Sakaki:2013bfa,Calibbi:2015kma,Freytsis:2015qca, Fortes:2015jaa,Kim:2015zla,Gripaios:2015gra, Bhattacharya:2015ida, Bauer:2015knc,Hati:2015awg,Fajfer:2015ycq,Barbieri:2015yvd,Cline:2015lqp,Dorsner:2016wpm,Dumont:2016xpj, Boucenna:2016wpr, Das:2016vkr,Nandi:2016wlp,Li:2016vvp,Feruglio:2016gvd} have examined possible beyond SM (BSM) origins for this anomaly. In many cases NP not only affects the $\bddstn$ rates compared to SM expectations, but also modifies the differential phase space distributions of the $\bddstn$ process.  Many studies have examined possible changes in the $q^2 \equiv (p_B - p_{\dds})^2$ invariant mass distribution, in order to assess the viability of NP models. An advantage of this observable, which is measured to moderate precision~\cite{Lees:2013uzd}, is that interference effects arising from decays of the $\tau$ and the $D^*$ are absent in $d\Gamma/dq^2$, provided there are no phase space cuts. In this case, one can treat the $\tau$ and $D^*$ as stable particles in the $b \to c \tau \tn$ decay.

The experimental measurements of $R(\dds)$ and other observables are, however, complicated by several considerations. First, prompt decay of both the $\tau$ and $D^*$ means that the $\tau$ and $D^*$ themselves are not external states. The non-negligible $\tau$ mass opens up significant contributions from both $\tau$ spin states, so that the consequent $\tau$ interference effects can be formally of order $m_\tau/m_B$ in the SM.  Moreover, SM--NP interference that is chirally suppressed by $m_\tau/m_B$ when treating the $\tau$ as stable, can become $\mathcal{O}(1)$ once interference between $\tau$ spin states is included.  Interference effects among the $D^*$ spin states are typically always $\mathcal{O}(1)$. Second, the presence of multiple neutrinos in the final state reduces the overall number of experimentally accessible observables, preventing full reconstruction of the underlying $\bddstn$ event. Once the full $\tau$ and $D^*$ decay phase space is considered, which contains at least five final-state particles, kinematic observables other than $q^2$ become available to probe the NP structure, e.g., the charged lepton energy, $E_\ell$, or the $\pi$--$\ell$ opening angle. Kinematic distributions of such observables are sensitive to these $\tau$ and $D^*$ interference effects, as are their expectation values integrated over the full phase space. Third, experimentally unavoidable phase space cuts, including both missing mass and lepton momentum cuts used to reduce backgrounds, imply that interference effects between the $\tau$ and $D^*$ spin states affect all pertinent measurements, including $d \Gamma/dq^2$. The experimental acceptances in the presence of NP may therefore differ from the SM ones used to extract $R(\dds)$. 

To properly capture all these effects, one must compute the matrix elements for the full $B \to D\tau(\to X \tnbar) \tn$ and $B \to D^*( \to D Y)\tau(\to X \tnbar) \tn$ processes, treating both the $\tau$ and $D^*$ as internal states. Computations of the corresponding full matrix elements for the SM only have long been available and implemented in prevalently used Monte Carlo generators, such as \texttt{EvtGen}~\cite{Lange:2001uf, Ryd:2005zz}. Computations for various parts of the full processes with NP are also available~\cite{Tanaka:2012nw, Duraisamy:2013kcw, Hagiwara:2014tsa, Becirevic:2016hea, Alok:2016qyh, Alonso:2016gym, Bordone:2016tex,Ivanov:2016qtw}, variously omitting the coherent $D^*$ decays and interference effects, the $\tau$ decays and interference effects, the NP interference effects with the SM, or combinations thereof. In this work, we present a set of generalized NP helicity amplitudes, i.e., matrix elements carrying explicit quantum numbers and full differential phase space dependence, for the full $B \to \dds (\to D Y) \tau (\to X\tnbar)\tn$ processes, in particular for $X = \ell \nu$ or $\pi$ and $Y = \pi$ or $\g$.  We contemplate NP arising from all possible four-Fermi operators with $\bar{b} c\, \bar{\nu} \tau$ flavor structure.  We include possible $CP$ violating NP, which may introduce additional large interference effects, and right-handed neutrinos, should they be Dirac. (Some of these operators may also be constrained by other flavor-diagonal and flavor-changing processes in the neutrino sector, but the current limits do not significantly constrain the scale of these operators beyond what is probed in $\bddstn$.) As such, this paper may be considered as an extension of Ref.~\cite{Goldberger:1999yh} to include all the effects mentioned above.

In practice, experiments measure $R(\dds)$ via a simultaneous fit of the expected signal distribution plus irreducible backgrounds, where the normalizations of various background components are allowed to vary. Including NP contributions in this fit requires estimation of the efficiencies and acceptances for the SM+NP signal via Monte Carlo (MC) simulations. Given the level of accuracy required by the anticipated high luminosity future of both LHCb and Belle~II, the MC datasets become impractically large once detector simulations are included.  In order to explore and run fits over the full space of BSM scenarios within reasonable timescales, one requires an efficient means to compute event weights, with which the fully simulated MC sample can be reweighted. With judicious choices of spinor phase and basis conventions and phase space coordinates, the helicity amplitudes for the $B \to D\tau(\to X \tnbar) \tn$ and $B \to D^*( \to D Y)\tau(\to X \tnbar) \tn$ processes can be expressed explicitly and compactly. Such explicit and compact expressions allow for very efficient computation of the relevant matrix elements required for reweighting the MC samples: The number of terms in the amplitude-level computation scales linearly as $\mathcal{O}(\sum_n m_n )$ for the inclusion of $n$ NP currents, each with $m_n$ internal quantum numbers, compared to $\mathcal{O}\big((\sum_n m_n)^2\big)$ for approaches that calculate the matrix element squared directly. A software package implementing these results, for use by experimental collaborations, is under preparation.\footnote{\texttt{Hammer}: Helicity Amplitude Module for Matrix Element Reweighting~\cite{Hammer:2016}.} 

In Sec.~\ref{sec:defs} we establish our notation and conventions. After deriving the amplitudes in Sec.~\ref{sec:HA}, we proceed to consider example applications of this efficient computational construction. We construct a MC method in Sec.~\ref{sec:apps}, in which MC data samples are reweighted with matrices of weights. This reweighting need only be
performed once per sample, and the result can be used to generate data for any new physics model. Post-reweighting, for any set of NP four-Fermi couplings, the distributions of kinematic observables $\mathcal{O}_i$ in $b_{i}$ bins can be generated by a smaller set of only $\sum_i b_i$ linear operations. The general problem of reweighting a large MC dataset between different NP theories is thereby reduced to a much smaller set of linear operations. We use this strategy to efficiently generate 1D and 2D distributions in ten kinematic observables, including lepton and pion energies and opening angles, with and without phase space cuts, over a range of NP couplings. To demonstrate the usefulness of efficiently producing multidimensional distributions, we present a sample bivariate analysis that exhibits higher distinguishing power between SM and NP theories, compared to using only single kinematic~distributions.

\section{Construction}
\label{sec:defs}

\subsection{Operator basis}

In addition to the SM four-Fermi interaction, we consider a complete set of four-Fermi NP operators mediating $\bbar \to \cbar \tau^+\tn$ decay, choosing an operator basis
\begin{subequations}
\label{eqn:FFOD}
\begin{align}
	\text{Vector:}~ & \phantom{-}i2\sqrt{2}V_{cb}G_F \bigg(\frac{m_W}{\Lambda_{V}}\bigg)^2\, \Big[\bbar\big(\alVL \g^\mu P_L + \alVR \g^\mu P_R\big)c\Big] \Big[\tnbar\big(\beVL \g_\mu P_L + \beVR \g_\mu P_R\big) \tau \Big]\,, \\[2pt]
	\text{Scalar:}~ & -i2\sqrt{2}V_{cb}G_F \bigg(\frac{m_W}{\Lambda_{S}}\bigg)^2\, \Big[\bbar\big(\alSL P_L + \alSR P_R\big)c\Big] \Big[\tnbar\big(\beSL P_R + \beSR P_L\big) \tau\Big]\,, \\[2pt]
	\text{Tensor:}~ & -i2\sqrt{2}V_{cb}G_F \bigg(\frac{m_W}{\Lambda_T}\bigg)^2\, \Big\{\Big[\bbar\big( \alTR \sigma^\mn P_R\big) c\Big] \Big[\tnbar\big( \beTL \sigma_\mn P_R \big) \tau \Big] \notag\\*
	& \qquad\qquad\qquad\qquad\qquad + \Big[\bbar\big( \alTL \sigma^\mn P_L\big) c\Big]\Big[\tnbar\big( \beTR \sigma_\mn P_L \big) \tau \Big] \Big\}\,.
\end{align}
\end{subequations}
Here we have classified each operator according to the Lorentz structure -- scalar, vector, or tensor -- of the contracted quark and lepton currents, $\bbar \Gamma c$ and $\tnbar \Gamma \tau$.  The CP conjugate operators for $b \to c \tau^-\tnbar$ are obtained by complex conjugation. (We are careful to label the tau neutrino in  $\bbar \to \cbar \tau^+ \tn$  distinctly from the tau antineutrino in $\tau \to \tnbar X$, and from the light lepton flavored neutrino for $X = \ell \nu_\ell$. Henceforth we drop all other bars and sign superscripts where the meaning is unambiguous.) We use the convention $\sigma^{\mn} \equiv (i/2)[\g^\mu,\g^\nu]$. 

NP couplings to the quark and lepton currents are denoted by $\alpha$ and $\beta$, respectively, normalized to $g_{2}V_{cb}/\sqrt{2}$ and $g_{2}/\sqrt{2}$, where $g_2$ is the $SU(2)$ electroweak coupling and $V_{cb}$ is the usual CKM element, while the scale of the operator is normalized to the $W$ mass, $m_W$. If one views each operator as a tree-level exchange of a fictitious particle, then $\alpha$ and $\beta$ correspond to its quark and lepton current couplings, respectively, and $\Lambda_{S,V,T}$ corresponds to the mediator mass. The NP couplings may be complex in general, admitting multiple sources of $CP$ violation. We label the chirality of the leptonic $\beta$ couplings according to the tau neutrino chirality, in order to easily distinguish between contributions involving left- and right-handed neutrinos, and hence contributions that do or do not interfere with the SM operator. Neglecting neutrino masses, $\beta_L$ and $\beta_R$ terms do not interfere. The chirality of the quark couplings $\alpha_{L,R}$ are defined by the chirality of the charm quark.  The identity 
\begin{equation}
	\label{eqn:SMNI}
	\sigma^\mn \g^5 \equiv \frac{i}{2}\, \epsilon^{\mn \rho \sigma} \sigma_{\rho \sigma}\,,
\end{equation}
with $\epsilon^{0123} = +1$,\footnote{Our sign conventions imply that $\text{Tr}[\g^\mu\g^\nu\g^\sigma\g^\rho\g^5] = -4i \epsilon^{\mu\nu\rho\sigma}$. Fixing instead sign conventions such that $\text{Tr}[\g^\mu\g^\nu\g^\sigma\g^\rho\g^5] = +4i \epsilon^{\mu\nu\rho\sigma}$, as done in many places in the literature, changes the sign of eq.~\eqref{eqn:SMNI}, as well as the sign of $g(q^2)$ and $a_{T_{\pm,0}}(q^2)$ in eqs.~\eqref{eqn:HQETDS}.} guarantees the absence of $\alTL\beTL$ or $\alTR\beTR$ terms, so that there are only two tensor operators. This yields a total of ten independent four-Fermi NP operators. Neutrino flavor-violating effects are GIM-suppressed and may be neglected. Finally, we assume in this paper that $\tau$ decays are described by the SM, supported by the good agreement of SM predictions with $\tau$ decay data~\cite{Agashe:2014kda}.  

\subsection{Form factors}
Lorentz symmetry ensures that for the $B \to \dds$ transitions, the scalar, pseudoscalar, vector, axial vector and tensor currents have one (zero), zero (one), two (one), zero (three) and one (three) independent form factors, respectively. We define
\begin{equation}
	q^\mu \equiv p^\mu_B - p^\mu_{\dds}\,, 
\end{equation}
so that $q^2$ is the only unfixed Lorentz invariant in the $B \to \dds$ decay. Note $m_\tau^2 \leq q^2 \leq (m_B - m_{\dds})^2$, and that $q^\mu$ is equivalently the momentum flowing to the $\tau \tn$ pair. For $\bar{B} \to D$ we adopt the following conventions and definitions for the form factors, 
\begin{subequations}
\begin{align}
	 \ampBb{D}{\cbar\,b} & \equiv f_S(q^2) \,, \\
	 \ampBb{D}{\cbar \g^\mu b} & \equiv f_+(q^2) (p_B + p_D)^\mu + [f_0(q^2) - f_+(q^2)]\, \frac{m_B^2 - m_D^2}{q^2}\, q ^\mu\,, \\
	 \ampBb{D}{\cbar \sigma^\mn b} & \equiv i f_T(q^2) \Big[(p_B + p_D)^{\mu}q^{\nu} - (p_B + p_D)^{\nu}q^{\mu}\Big]\,.
\end{align}
\end{subequations}
The pseudoscalar and axial vector currents $\ampBbs{D}{\cbar \g^5 b} \equiv 0$ and $\ampBbs{D}{\cbar \g^\mu \g^5 b} \equiv 0$, while the axial tensor current $\ampBbs{D}{\cbar \sigma^\mn \g^5 b}$ is fixed by the identity \eqref{eqn:SMNI}.
Under these conventions, at leading order in $\lqcd/m_{b,c}$, these form factors are
\begin{subequations}
\begin{align}
	f_S(q^2) & = \xi(w)\, \frac{(m_B + m_D)^2 - q^2}{2 \sqrt{m_D m_B}}\,, \\
	f_+(q^2) & = \xi(w)\, \frac{m_B+m_D}{2 \sqrt{m_D m_B}}\,, \\
	f_0(q^2) & = \xi(w)\, \frac{(m_B + m_D)^2 - q^2}{2 \sqrt{m_D m_B}\, (m_B + m_D)}\,, \\
	f_T(q^2) & = \frac{\xi(w)}{2 \sqrt{m_D m_B}}\,,
\end{align}
\end{subequations}
where $\xi(w)$ is the Isgur-Wise function~\cite{Isgur:1989vq, Isgur:1989ed}. These relations are understood for the value of the recoil parameter $w \equiv v_B \cdot v_{D^{(*)}}= (m_B^2 + m_{D^{(*)}}^2 - q^2) / (2m_B m_{D^{(*)}})$. Under $CP$ conjugation, the form factors for the conjugate $B \to \bar{D}$ process are
\begin{subequations}
\label{eqn:FFD}
\begin{align}
	 \ampB{\Dbar}{\bbar\,c} & = f_S(q^2)\,, \\
	 \ampB{\Dbar}{\bbar \g^\mu c} & = -f_+(q^2)\, (p_B + p_D)^\mu - \big[f_0(q^2) - f_+(q^2)\big]\, \frac{m_B^2 - m_D^2}{q^2}\, q ^\mu\,, \\
	 \ampB{\Dbar}{\bbar \sigma^\mn c} & = -i f_T(q^2) \Big[(p_B + p_D)^{\mu}q^{\nu} - (p_B + p_D)^{\nu}q^{\mu}\Big]\,, 
\end{align}
\end{subequations}
noting in particular the sign change for the tensor and vector currents.

Similarly for $\Bbar \to D^*$ we define
\begin{subequations}
\begin{align}
	 \ampBb{D^*}{\cbar \g^5 b} & \equiv a_0(q^2)\dotpr{\varepsilon^*}{p_B} \,,\\
	 \ampBb{D^*}{\cbar \g^\mu b} & \equiv -i g(q^2)\, \epsilon^{\mn \rho \sigma}\, \varepsilon^*_\nu\, (p_B + p_{D^*})_\rho\, q_\sigma\,, \\
	 \ampBb{D^*}{\cbar \g^\mu \g^5 b} & \equiv  {\varepsilon^*}^\mu f(q^2) + a_{+}(q^2) \dotpr{\varepsilon^*}{p_B}(p_B + p_{D^*})^\mu + a_{-}(q^2) \dotpr{\varepsilon^*}{p_B}q^\mu\,,  \\
	 \ampBb{D^*}{\cbar \sigma^\mn b} & \equiv -a_{T_+}(q^2)\, \epsilon^{\mn \rho \sigma} \varepsilon^*_\rho (p_B + p_{D^*})_\sigma - a_{T_-}(q^2)\, \epsilon^{\mn \rho \sigma} \varepsilon^*_\rho\, q_\sigma \notag \\*
	 & \quad - a_{T_0}(q^2) \dotpr{\varepsilon^*}{p_B} \epsilon^{\mn \rho \sigma} (p_B + p_{D^*})_\rho\, q_\sigma\,.
\end{align}
\end{subequations} 
The matrix element of the scalar current vanishes, $\ampBbs{D^*}{\cbar\, b} \equiv 0$, while the axial tensor current matrix element $\ampBbs{D^*}{\cbar \sigma^\mn \g^5 b}$ is fixed by the identity \eqref{eqn:SMNI}. At leading order in $\lqcd/m_{b,c}$, these form factors are
\begin{subequations}
\label{eqn:HQETDS}
\begin{align}
	a_0(q^2) & = \xi(w)\, \sqrt{\frac{m_{D^*}}{m_B}}\,,\\
	a_+(q^2) & = -a_-(q^2) = -g(q^2) =    \frac{\xi(w)}{2\sqrt{m_{D^*}m_B}}\,, \\
	f(q^2) & = -\xi(w)\, \frac{(m_B + m_{D^*})^2 - q^2}{2 \sqrt{m_{D^*}m_B}}\,,\\
	a_{T_\pm}(q^2) & = \pm \xi(w)\, \frac{m_B \pm m_{D^*}}{2 \sqrt{m_{D^*}m_B}}\,,\\
	a_{T_0}(q^2) & = 0\,.
\end{align}
\end{subequations}
Under $CP$ conjugation, the form factors for the conjugate $B \to \Dbar^*$ process are
\begin{subequations}
\label{eqn:FFDS}
\begin{align}
	 \ampB{\Dbar^*}{\bbar \g^5 c} & = a_0(q^2)\dotpr{\varepsilon^*}{p_B} \,,\\[4pt] 
	 \ampB{\Dbar^*}{\bbar \g^\mu c} & = i g(q^2)\, \epsilon^{\mn \rho \sigma} \varepsilon^*_\nu (p_B + p_{D^*})_\rho\, q_\sigma\,, \\
	 \ampB{\Dbar^*}{\bbar \g^\mu \g^5 c} & =  {\varepsilon^*}^\mu f(q^2) + a_{+}(q^2) \dotpr{\varepsilon^*}{p_B}(p_B + p_{D^*})^\mu  + a_{-}(q^2) \dotpr{\varepsilon^*}{p_B}q^\mu\,, \\
	 \ampB{\Dbar^*}{\bbar \sigma^\mn c} & = a_{T_+}(q^2)\, \epsilon^{\mn \rho \sigma} \varepsilon^*_\rho (p_B + p_{D^*})_\sigma + a_{T_-}(q^2)\, \epsilon^{\mn \rho \sigma} \varepsilon^*_\rho\, q_\sigma \notag \\
	 & \qquad + a_{T_0}(q^2) \dotpr{\varepsilon^*}{p_B}\, \epsilon^{\mn \rho \sigma} (p_B + p_{D^*})_\rho\, q_\sigma\,, 
\end{align}
\end{subequations}
noting that the pseudoscalar and axial currents do not change sign.

\subsection{Helicity angles}
The helicity amplitudes are most simply expressed in terms of the $(\theta,\phi)$ helicity angles for each vertex of the $B \to \ddsbar (\to \Dbar Y)\tau^+(\to X \tnbar)\tn$ amplitude.\footnote{Helicity angles and momenta are labelled according to the $\bbar \to \cbar$ process. Corresponding definitions for the conjugate process follow by replacing all particle labels with their antiparticles.} That is, we factorize the phase space of the process into a series of rest frames in the (off-shell) cascade $B \to \dds (\to DY) \Wbar (\to  \tn \tau ( \to  \tnbar W (\to X)))$ and so on. Here, for the purpose of defining helicity angles, we treat the $\tau \tn$ pair as originating from a fictitious $\Wbar$ particle in the $B \to \dds$ transition, with momentum $q^\mu$. Similarly we define $p^\mu$ to be the momentum of the $W^*$ in the $\tau$ decay, and $p^2 \in [0,m_\tau^2]$ neglecting the daughter charged lepton's mass. (Hereafter we always label the momenta of massive particles with the base symbol $p$ and those of massless particles with the base symbol $k$.)

\begin{figure}[t]
	\centering\includegraphics[width=12cm]{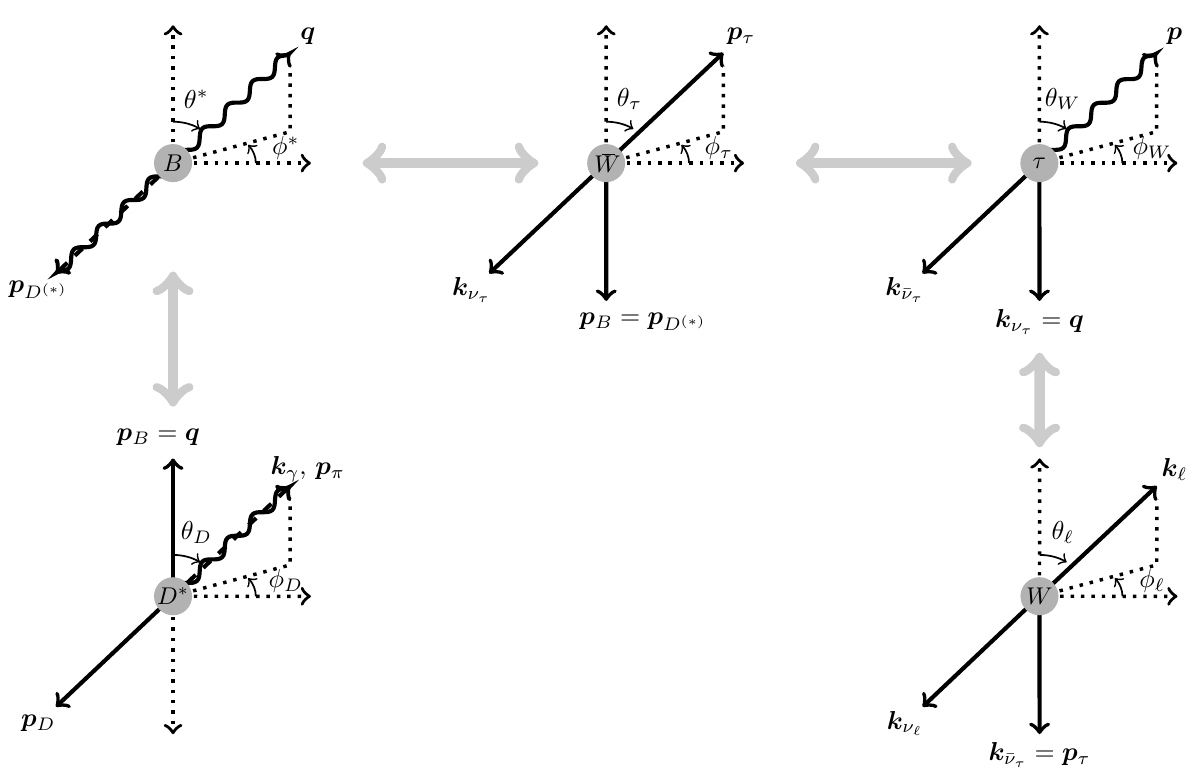}
	\caption{Helicity angle definitions with respect to spatial momenta (bold symbols) in the sequence of particle rest frames. Each subfigure is drawn in the rest frame of the particle denoted in the central grey disk. Transformations between frames are achieved by Euler rotations and Lorentz boosts, denoted by gray arrows (see text for details).} 
\label{fig:HASD}
\end{figure}

In Fig.~\ref{fig:HASD} we show schematically the helicity angle definitions for $B \to \dds (\to DY) \Wbar (\to  \tn \tau ( \to  \tnbar W (\to \ell\nu_\ell)))$, with $Y = \pi$ or $\g$. Explicit expressions for these helicity angles in terms of Lorentz invariant objects are provided in Appendix~\ref{app:HAE}.  The polar $\theta$ angles in Fig.~\ref{fig:HASD} are well-defined rest frame by rest frame. The orientation of the azimuthal $\phi$ angles is, however, defined with respect to an arbitrary direction in the $B$ rest frame, $(\theta^*, \phi^*)$, combined with a sequence of parent-daughter frame transformations. As the $B$ is a spin-$0$ state, the $(\theta^*, \phi^*)$ angles themselves are unphysical, and vanish from all amplitudes, but we nonetheless keep these angles explicit in Fig.~\ref{fig:HASD}. In a parent rest frame with daughter polar coordinates $(\theta,\phi)$, the parent-daughter frame transformation is defined to be the sequential $z\, y' z''$ Euler rotations $R_{z''}(\phi) R_{y'}(-\theta) R_z(-\phi)$, followed by a Lorentz boost along the $z''$ axis to the daughter frame. These Euler rotations transform to a frame in which the daughter momentum is aligned with the $z''$ axis, while preserving a line of nodes orthogonal to the plane of the daughter momentum and $z$ axis. These conventions ensure that apart from the polar $\theta$ angles, only the relative twist angles $\phtau-\phW$, $\phell - \phW$ and $\phD - \phtau$ are physical. 
	
\subsection{Phase space}

The phase space integration limits are $[0,\pi)$ and $[0,2\pi)$ for each polar and azimuthal helicity angle. In these coordinates, the full phase space measure can be straightforwardly factorized into $\bddstn$, $\tau \to X \tnbar$ and $D^* \to D\pi$, $D\g$ pieces. These are
\begin{align}
	 d\PS_{\bddstn}   & = \frac{1}{1024 \pi^5} \bigg(1 - \frac{m_\tau^2}{q^2}\bigg)\frac{\Pw}{m_B}\, d\Omega_\tau d\Omega^* dq^2\,, \notag\\
	 d\PS_{\tau \to \ell \nu_\ell \tnbar}  & = \frac{1}{2048 \pi^5}\bigg(1 - \frac{p^2}{m_\tau^2} \bigg) d\Omega_\ell d\Omega_W dp^2\,, \qquad \label{eqn:PSM}
\end{align}
while $d\PS_{\tau \to \pi \tnbar} = (1 - m_\pi^2/m_\tau^2)/(32 \pi^2)\,d\Omega_\pi$, $d\PS_{D^* \to D\pi} = \Pp/(16 \pi^2 m_{D^*})\,d\Omega_D$ and $d\PS_{D^* \to D\g} = [1 - m_D^2/m_{D^*}^2]/(32 \pi^2)\,d\Omega_D$. Here the spatial momentum of the $\tau \tn$ pair in the $B$ rest frame and of the pion in the $D^*$ rest frame are, respectively,
\begin{align}
	\Pw & = \frac{m_B}2\, \lambda\big(m_{\dds}/m_B,\ \sqrt{q^2}/m_B\big)\,, \notag \\
	\Pp  & = \frac{m_{D^*}}2\, \lambda\big(m_D/m_{D^*},\ m_\pi/m_{D^*}\big)\,,
\end{align}
with $\lambda(x,y) \equiv \sqrt{[1 - (x-y)^2][1- (x+y)^2]}$ the usual phase space factor.

\section{Amplitudes}
\label{sec:HA}

The helicity amplitudes for the full $B \to \dds(\to D Y)\tau(\to X \tnbar) \tn$ process carry only quantum numbers of external particles (i.e., not the $\tau$ and $D^*$ spins) corresponding to certain convenient basis choices for external spinors and polarization vectors. For $X = \ell \nu_\ell$, these are the spins $s_{\tn}$, $s_{\tnbar}$, $s_\ell$, $s_{\nu_\ell} = \dn$, $\up$ that label the helicity amplitudes below, and also the photon helicity $\kappa = \pm$ in the case of $D^* \to D\g$.

The azimuthal helicity angles arise as phases in the helicity amplitudes. These phases are odd under $CP$, along with those that occur in the NP $\alpha$ or $\beta$ couplings. In the remainder of this paper, we shall consider explicit expressions for only the $\bbar \to \cbar$ process.  Results for the $CP$ conjugate $b \to c$ process are obtained by conjugation of all these phases, i.e.,
\begin{equation}
	\mathcal{A}_{b \to c}^s(\theta,\phi; \alpha, \beta) = \mathcal{A}_{\bbar \to \cbar}^{\bar{s}}(\theta, -\phi; \alpha^*, \beta^*)\,,
\end{equation}
where $s$ is the set of quantum numbers of all external states, and $\bar{s}$ the corresponding $CP$ conjugate, obtained by interchanging all spins and helicities with their conjugates.

Since we assume that $\tau$ decays are described by the SM, and we can neglect the light charged daughter lepton mass, it is always the case that $s_{\tnbar} = \up$, $s_\ell = \up$, and $s_{\nu_\ell} =\dn$, such that our choice of spinor basis for massless states coincides with the usual helicity basis. We drop these quantum numbers from the amplitude labelling below, with the understanding that all other amplitudes are zero. For the SM, $s_{\tn} = \dn$ only. However, in the presence of NP currents involving left-(right-)handed $\nu_\tau$, associated with $\beta_L$ ($\beta_R$) couplings, one may further have $s_{\tn} = \dn$ ($s_{\tn} = \up$) contributions that do (do not) interfere with the SM.

\subsection{Amplitude factorization and \texorpdfstring{$\tau$}{T} spinor basis}

It is convenient to express the helicity amplitudes factorized into $B \to \dds(\to DY)\tau \tn$ and $\tau \to X\tnbar$ pieces, not only for the sake of presentation, but also in order to enable the $B \to \dds(\to DY)\tau \tn$ results to be used modularly with respect to different choices of $\tau \to X\tnbar$. To obtain the square of the polarized matrix elements, one sums over the internal $\tau$ spin, $s_\tau = 1,\,2$,\footnote{For a massive fermion, we label spin states by $1$ and $2$ (see, e.g., p.48 in Ref.~\cite{Peskin:1995ev}).} before squaring,
\begin{equation}
	\label{eqn:AF}
	|\mathcal{M}|^2_{B \to D\tau(\to X \tnbar) \tn}  = \sum_{s_{\tn}, s_{\tnbar}, s_X, s_Y} \Big| \sum_{s_\tau} [\mathcal{A}_{\bdtn}]^{s_{\tn}}_{s_{\tau}} [\mathcal{A}_{\tau \to X \tnbar}]^{s_{\tnbar} s_X}_{s_\tau} \Big|^2 ,
\end{equation}
and similarly for $B \to D^*( \to D Y)\tau(\to X \tnbar) \tn$. Here $s_{X}$ ($s_Y$) is the set of quantum numbers of the $X$ ($Y$) external state: $s_X = \{ s_\ell, s_{\nu_\ell} \}$ for $X = \ell \nu_\ell$, $s_Y = \kappa$ for $Y = \g$ and $s_{X}$ ($s_Y$) is empty for $X = \pi$ ($Y = \pi$). The fully differential decay rates are then
\begin{align}
	d\Gamma_{B \to D\tau(\to X \tnbar) \tn} & = \frac1{2m_B}\, \frac1{2m_\tau \Gamma_\tau}\, |\mathcal{M}|^2_{B \to D\tau(\to X \tnbar) \tn}  d\PS_{\bdtn}d\PS_{\tau \to X \tnbar}\,, \\
	d\Gamma_{B \to D^*( \to D Y)\tau(\to X \tnbar) \tn} & = \frac1{2m_B}\, \frac1{2m_\tau \Gamma_\tau}\, \frac1{2m_{D^*}\, \Gamma_{D^*}}\, |\mathcal{M}|^2_{B \to D^*( \to D Y)\tau(\to X \tnbar) \tn} \notag\\*
	& \qquad \qquad \times d\PS_{\bdstn}d\PS_{\tau \to X \tnbar}d \PS_{D^* \to D Y}\,,
\end{align}
where we have included the factorized phase space measures~\eqref{eqn:PSM} as well as $\tau$ and $D^*$ propagators, using the narrow width approximation for both states. 

In order to permit extension of the results below to any $\tau \to X\tn$ decay, we specify here our choice for the $\tau$ spinor basis and phase conventions. Calculation of the helicity amplitudes are achieved by decomposing momenta and spinors (or polarizations) of massive states onto a lightcone basis. For the $\tau$, we choose the $\nu_\tau$ momentum $k_{\tn}$ as a null reference momentum. In the $\tau$ rest frame, using phase space coordinates as defined in Fig.~\ref{fig:HASD}, the Dirac spinor basis for the $\tau^+$ is
\begin{equation}
	\label{eqn:TSBC}
	\bar{v}^1(p_\tau; k_{\tn}) = h_1(s_{\tn}) \big( \sqrt{m_\tau}\,,\, 0\,,\, 0\,,\, -\sqrt{m_\tau} \big)\,,\qquad \bar{v}^2(p_\tau; k_{\tn}) = h_2(s_{\tn}) \big( 0\,,\, \sqrt{m_\tau}\,,\, \sqrt{m_\tau}\,,\, 0 \big)\,,
\end{equation}
for $\g^\mu$ in the Dirac basis and $\g^5 = \text{diag}\{-\bm{1}_2,\bm{1}_2\}$, $P_{R,L} \equiv (1 \pm \g^5)/2$. While the factorization~\eqref{eqn:AF} permits modularity under choices of $\tau \to X\tn$, it may also introduce unphysical manifestations of the azimuthal helicity angle $\phtau$ in each amplitude factor, which disappear under summation over $s_\tau$. It is, however, far more computationally efficient to permit only physical phases --- the relative azimuthal twist angles --- to appear in each helicity amplitude factor. To ensure that $\phtau$ appears only in the physical combinations $\phD - \phtau$  and $\phtau - \phW$ in the $B \to \dds(\to DY)\tau \tn$ and $\tau \to X\tn$ helicity amplitudes, respectively, we introduced in eq.~\eqref{eqn:TSBC} an additional spinor phase function, $h_{s_\tau}(s_{\tn})$, defined with respect to $s_{\tn}$, such that
\begin{equation}
	h_{1}(\dn) = 1 = h_{2}(\up)\,,\qquad h_{1}(\up) = e^{i\phtau}\,\qquad h_{2}(\dn) = e^{-i\phtau}\,.
\end{equation}
This additional phase factor in the $\tau^+$ spinors is balanced by a cancelling phase factor $e^{\pm i\phtau}$ in the corresponding $B \to \dds(\to DY)\tau \tn$ amplitudes. We emphasize that this is merely a bookkeeping device, that does not affect the physical phase structure of the full $B \to D^*( \to D Y)\tau(\to X \tnbar) \tn$ helicity amplitudes. Under this phase convention the $\tau \to X\tnbar$ helicity amplitudes therefore carry $s_{\tn}$ as a quantum number, even though $\tn$ itself is not involved in the $\tau$ decay. 

The quantum numbers in eq.~\eqref{eqn:TSBC} need only be matched with those in each of the $B \to \dds(\to DY)\tau \tn$ helicity amplitudes below to identify the corresponding $\tau$ spinor and phase to be used to compute the $\tau$ decay helicity amplitude of interest. We provide below explicit expressions for the $\tau \to \ell \nu_\ell \tnbar$ and $\tau \to \pi \tnbar$ amplitudes under these conventions.

\subsection{\texorpdfstring{$\bdtn$}{bdtn}}

Let us now proceed to present the helicity amplitudes. For readability, we group terms by form factors. For $\bdtn$, the helicity amplitudes $[\mathcal{A}_{\bdtn}]^{s_{\tn}}_{s_{\tau}} \equiv \mathcal{A}^{s_{\tn}}_{s_\tau}$ are
\begin{subequations}
\begin{align}
\mathcal{A}^\dn_1 = -i2 \sqrt{2} V_{cb} G_F \sqrt{q^2 - m_{\tau}^2}\, \bigg\{
    & \frac{1}{2} f_S(q^2)  (\alSL +\alSR ) \beSL  r_{S}^2 \notag \\*
    & +\frac{f_0(q^2) (-m_B^2 + m_D^2) m_{\tau} (1 + (\alVL + \alVR)\beVL r_V^2)}{2 q^2 } \notag \\*
    & +\frac{f_+(q^2)  m_{B} m_{\tau} \Pw  (1 + (\alVL + \alVR)\beVL r_V^2) \cos(\thtau)}{q^2 } \notag \\*
    & -4 f_T(q^2) m_{B} \Pw  \alTR  \beTL  r_{T}^2 \cos(\thtau)\bigg\}\,, \\[8pt]
\mathcal{A}^\dn_2 = -i2 \sqrt{2} V_{cb} G_F \sqrt{q^2 - m_{\tau}^2}\, \bigg\{
    & -\frac{f_+(q^2)  m_{B} \Pw  (1 + (\alVL + \alVR)\beVL r_V^2) \sin(\thtau)}{\sqrt{q^2 }} \notag \\*
    & +\frac{4 f_T(q^2)  m_{B} m_{\tau} \Pw  \alTR  \beTL  r_{T}^2 \sin(\thtau)}{\sqrt{q^2 }}\bigg\}\,, \\[8pt]
\mathcal{A}^\up_1 = -i2 \sqrt{2} V_{cb} G_F \sqrt{q^2 - m_{\tau}^2}\, \bigg\{
    & -\frac{f_+(q^2) m_{B} \Pw  (\alVL +\alVR ) \beVR  r_{V}^2 \sin(\thtau)}{\sqrt{q^2 }} \notag \\*
    & +\frac{4 f_T(q^2) m_{B} m_{\tau} \Pw  \alTL  \beTR  r_{T}^2 \sin(\thtau)}{\sqrt{q^2 }}\bigg\}\,,\\[8pt]
\mathcal{A}^\up_2 =    -i2 \sqrt{2}  V_{cb} G_F \sqrt{q^2 - m_{\tau}^2}\, \bigg\{
    & -\frac{1}{2} f_S(q^2)  (\alSL +\alSR ) \beSR  r_{S}^2 \notag \\*
    & +\frac{f_0(q^2)  (m_B^2 - m_D^2) m_{\tau} (\alVL +\alVR ) \beVR  r_{V}^2}{2 q^2 } \notag \\*
    & -\frac{f_+(q^2)  m_{B} m_{\tau} \Pw (\alVL +\alVR ) \beVR  r_{V}^2 \cos(\thtau)}{q^2 } \notag \\*
    & +4 f_T(q^2) m_{B} \Pw \alTL \beTR r_{T}^2 \cos(\thtau)\bigg\}\,, 
\end{align}
\end{subequations}
where $r_{V,S,T} \equiv m_W/\Lambda_{V,S,T}$. 

Expressions for the SM helicity amplitudes may be read off taking all $\alpha$'s or all $\beta$'s to zero. These SM results numerically match the output of \texttt{EvtGen}. In the SM, only $\mathcal{A}^\dn_1$ and $\mathcal{A}^\dn_2$ are non-zero, and contain terms that are all linear or zeroth order in $m_\tau$, respectively. Interference effects arising from decay of the $s_\tau = 1,2$ spin states to the same final state therefore enter at $\mathcal{O}(m_\tau/m_B)$ in the SM. When treating the $\tau$ as stable, interference terms for operators that respectively couple to $\bar{\nu}_{\tau L} \tau_L$ and $\bar{\nu}_{\tau L}\tau_R$, such as the $f_S f_+$ term between the NP scalar and SM vector operators within $\mathcal{A}^\dn_1$, are chirally suppressed as expected, entering only at order $m_\tau/m_B$. However, interference between $\tau$ spin states can produce $\mathcal{O}(1)$ contributions to these terms, e.g. the $f_S f_+$ interference term between $\mathcal{A}^\dn_1$ and $\mathcal{A}^\dn_2$. Similar conclusions follow for $B \to D^*\tau \tn$, below.

\subsection{\texorpdfstring{$B \to D^{*}(\to D \pi)\tau \tn$}{bdsdptn}}

The decay $D^* \to D\pi$ proceeds through the operator $\hat{g}_\pi {D^*}^\mu( \pi \partial_\mu D - D \partial_\mu \pi)$, in which $\hat{g}_\pi$ is the phenomenological coupling
\begin{equation}
	\hat{g}_\pi = \bigg[\frac{6 \pi m_{D^*}^2 \Gamma(D^* \to D\pi)}{\Pp^3}\bigg]^{1/2}\,,
\end{equation}
with $\hat{g}_\pi = (m_{D^*}/f_\pi) g_\pi$ in the notation of Ref.~\cite{Manohar:2000dt}. We define the functions
\begin{subequations}
\begin{align}
	\DpiA_\pm & \equiv \sin\thD \bigg[\cos^2\frac{\thtau}{2}\, e^{- i \phDtau} \pm \sin^2\frac{\thtau}{2}\, e^{ i \phDtau}\bigg]\,, \\
	\DpiA_0 & \equiv \cos\thD \sin \thtau\,,\\
	\DpiB_+ & \equiv \cos \thD\cos\thtau\,,\\
	\DpiB_-^R & \equiv \sin\thD\sin\thtau\cos\phDtau\,,\\
	\DpiB_-^I & \equiv \sin \thD\sin\thtau \sin\phDtau\,,\\
	\DpiB_0 & \equiv \cos\thD\,.
\end{align}
\end{subequations}
Under our phase and spinor conventions, the $s_\tau = 2$ ($s_\tau = 1$) helicity amplitudes are linear combinations of the $\DpiA$ ($\DpiB$) functions exclusively. Each set of $\DpiA$ or $\DpiB$ functions is $L^2(\mathbb{C})$ orthogonal under integration over the angular phase space $d\Omega_D d \Omega_\tau$. The $\DpiA$ functions are orthogonal with respect to $\DpiB$ once one accounts for the additional $e^{\pm i \phtau}$ phase that must occur in the integration measure, in accordance with our $\tau$ spinor phase conventions~\eqref{eqn:TSBC}. (This phase is encoded in the $\tau \to X\tnbar$ amplitudes below.) This $\DpiA$\,--\,$\DpiB$ orthogonality corresponds to the absence of $\tau$ interference effects in the total rate under integration over the full angular phase space, i.e., no angular phase space cuts, as expected. 

The helicity amplitudes $[\mathcal{A}_{B \to D^{*}(\to D \pi)\tau \tn}]^{s_{\tn}}_{s_{\tau}} \equiv [\mathcal{A}^\pi]^{s_{\tn}}_{s_\tau}$ are found to be
\begin{subequations}
\begin{align}
& [\mathcal{A}^\pi]^\dn_1  = -i 2 \sqrt{2} \hat{g}_\pi V_{cb}  G_F \Pp \sqrt{q^2 - m_{\tau}^2}\, \bigg\{
    \frac{a_0(q^2)  m_{B} \Pw  (\alSL -\alSR ) \beSL  r_{S}^2 \DpiB_0 }{m_{D^*}} \\*
    & +f(q^2)  m_{\tau} (1 + (\alVL - \alVR)\beVL r_V^2) \bigg[\frac{(- m_B^2 + m_{D^*}^2 + q^2) \DpiB_+ }{2 m_{D^*} q^2 }+\frac{m_{B} \Pw  \DpiB_0 }{m_{D^*} q^2 }-\frac{\DpiB_-^R }{\sqrt{q^2 }}\bigg] \notag \\ 
    & +\frac{2 i g(q^2)  m_{B} m_{\tau} \Pw  (1 + (\alVL + \alVR)\beVL r_V^2) \DpiB_-^I }{\sqrt{q^2 }}
    +\frac{a_-(q^2)  m_{B} m_{\tau} \Pw  (1 + (\alVL - \alVR)\beVL r_V^2) \DpiB_0 }{m_{D^*}} \notag \\ 
    & -a_+(q^2)  m_{B} m_{\tau} \Pw  (1 + (\alVL - \alVR)\beVL r_V^2) \bigg[\frac{2 m_{B} \Pw  \DpiB_+ }{m_{D^*} q^2 }+\frac{(- m_B^2 + m_{D^*}^2) \DpiB_0 }{m_{D^*} q^2 }\bigg] \notag \\ 
    & +\frac{8 a_{T_0}(q^2)  m_{B}^2 \Pw ^2 \alTR  \beTL  r_{T}^2 \DpiB_+ }{m_{D^*}}
    -2 a_{T_-}(q^2)  \alTR  \beTL  r_{T}^2 \bigg[\frac{(- m_B^2 + m_{D^*}^2 + q^2) \DpiB_+ }{m_{D^*}}-2 \sqrt{q^2 } \DpiB_-^R \bigg] \notag \\ 
    & +2 a_{T_+}(q^2)  \alTR  \beTL  r_{T}^2 \bigg[\frac{(m_B^2 + 3m_{D^*}^2 - q^2) \DpiB_+ }{m_{D^*}}+\frac{2 (m_B^2 - m_{D^*}^2) \DpiB_-^R }{\sqrt{q^2 }}+\frac{4 i m_{B} \Pw  \DpiB_-^I }{\sqrt{q^2 }}\bigg]\bigg\} \notag\\[10pt]
& [\mathcal{A}^\pi]^\dn_2  = -i 2 \sqrt{2} \hat{g}_\pi V_{cb} G_F \Pp \sqrt{q^2 - m_{\tau}^2}\, \bigg\{ \\ 
    & +f(q^2)  (-1 + (\alVR - \alVL)\beVL r_V^2) \bigg[\frac{(- m_B^2 + m_{D^*}^2 + q^2) \DpiA_0 }{2 m_{D^*} \sqrt{q^2 }}+\DpiA_- \bigg] \notag \\ 
    & -2 g(q^2)  m_{B} \Pw  (1 + (\alVL + \alVR)\beVL r_V^2) \DpiA_+  
    +\frac{2 a_+(q^2)  m_{B}^2 \Pw ^2 (1 + (\alVL - \alVR)\beVL r_V^2) \DpiA_0 }{m_{D^*} \sqrt{q^2 }} \notag \\ 
    & -\frac{8 a_{T_0}(q^2)  m_{B}^2 m_{\tau} \Pw ^2 \alTR  \beTL  r_{T}^2 \DpiA_0 }{m_{D^*} \sqrt{q^2 }}
    +2 a_{T_-}(q^2)  m_{\tau} \alTR  \beTL  r_{T}^2 \bigg[\frac{(- m_B^2 + m_{D^*}^2 + q^2) \DpiA_0 }{m_{D^*} \sqrt{q^2 }}+2 \DpiA_- \bigg] \notag \\ 
    & -2 a_{T_+}(q^2)  m_{\tau} \alTR  \beTL  r_{T}^2 \bigg[\frac{4 m_{B} \Pw  \DpiA_+ }{q^2 }+\frac{(m_B^2 + 3m_{D^*}^2 - q^2) \DpiA_0 }{m_{D^*} \sqrt{q^2 }}-\frac{2 (m_B^2 - m_{D^*}^2) \DpiA_- }{q^2 }\bigg]\bigg\}\notag\\[15pt]
& [\mathcal{A}^\pi]^\up_1  = -i 2 \sqrt{2} \hat{g}_\pi V_{cb}  G_F \Pp  \sqrt{q^2 - m_{\tau}^2}\, \bigg\{\\
    & +f(q^2)  (-\alVL +\alVR ) \beVR  r_{V}^2 \bigg[\frac{(- m_B^2 + m_{D^*}^2 + q^2) \DpiA_0 }{2 m_{D^*} \sqrt{q^2 }}+\DpiA_-^* \bigg] \notag \\ 
    & +2 g(q^2)  m_{B} \Pw  (\alVL +\alVR ) \beVR  r_{V}^2 \DpiA_+^* 
    +\frac{2 a_+(q^2)  m_{B}^2 \Pw ^2 (\alVL -\alVR ) \beVR  r_{V}^2 \DpiA_0 }{m_{D^*} \sqrt{q^2 }} \notag \\ 
    & +\frac{8 a_{T_0}(q^2)  m_{B}^2 m_{\tau} \Pw ^2 \alTL  \beTR  r_{T}^2 \DpiA_0 }{m_{D^*} \sqrt{q^2 }}
    +2 a_{T_-}(q^2)  m_{\tau} \alTL  \beTR  r_{T}^2 \bigg[\frac{(m_B^2 - m_{D^*}^2 - q^2) \DpiA_0 }{m_{D^*} \sqrt{q^2 }}-2 \DpiA_-^* \bigg] \notag \\ 
    & +2 a_{T_+}(q^2)  m_{\tau} \alTL  \beTR  r_{T}^2 \bigg[\frac{(m_B^2 + 3m_{D^*}^2 - q^2) \DpiA_0 }{m_{D^*} \sqrt{q^2 }}+\frac{4 m_{B} \Pw  \DpiA_+^* }{q^2 }-\frac{2 (m_B^2 - m_{D^*}^2) \DpiA_-^* }{q^2 }\bigg]\bigg\}\notag\\[15pt]
& [\mathcal{A}^\pi]^\up_2  = -i 2 \sqrt{2} \hat{g}_\pi V_{cb} G_F \Pp \sqrt{q^2 - m_{\tau}^2}\bigg\{
    \frac{a_0(q^2) m_{B} \Pw (-\alSL +\alSR ) \beSR r_{S}^2 \DpiB_0 }{m_{D^*}} \\*
    & +f(q^2)  m_{\tau} (-\alVL +\alVR ) \beVR  r_{V}^2 \bigg[\frac{(- m_B^2 + m_{D^*}^2 + q^2) \DpiB_+ }{2 m_{D^*} q^2 }+\frac{m_{B} \Pw  \DpiB_0 }{m_{D^*} q^2 }-\frac{\DpiB_-^R }{\sqrt{q^2 }}\bigg] \notag \\ 
    & -\frac{2 i g(q^2)  m_{B} m_{\tau} \Pw  (\alVL +\alVR ) \beVR  r_{V}^2 \DpiB_-^I }{\sqrt{q^2 }} 
    +\frac{a_-(q^2)  m_{B} m_{\tau} \Pw  (-\alVL +\alVR ) \beVR  r_{V}^2 \DpiB_0 }{m_{D^*}} \notag \\ 
    & +a_+(q^2)  m_{B} m_{\tau} \Pw  (\alVL -\alVR ) \beVR  r_{V}^2 \bigg[\frac{2 m_{B} \Pw  \DpiB_+ }{m_{D^*} q^2 }+\frac{(- m_B^2 + m_{D^*}^2) \DpiB_0 }{m_{D^*} q^2 }\bigg] \notag \\ 
    & +\frac{8 a_{T_0}(q^2)  m_{B}^2 \Pw ^2 \alTL  \beTR  r_{T}^2 \DpiB_+ }{m_{D^*}}
    -2 a_{T_-}(q^2)  \alTL  \beTR  r_{T}^2 \bigg[\frac{(- m_B^2 + m_{D^*}^2 + q^2) \DpiB_+ }{m_{D^*}}-2 \sqrt{q^2 } \DpiB_-^R \bigg] \notag \\ 
    & +2 a_{T_+}(q^2)  \alTL  \beTR  r_{T}^2 \bigg[\frac{(m_B^2 + 3m_{D^*}^2 - q^2) \DpiB_+ }{m_{D^*}}+\frac{2 (m_B^2 - m_{D^*}^2) \DpiB_-^R }{\sqrt{q^2 }}-\frac{4 i m_{B} \Pw  \DpiB_-^I }{\sqrt{q^2 }}\bigg]\bigg\}\,,\notag
\end{align}
\end{subequations}
where again $r_{V,S,T} \equiv m_W/\Lambda_{V,S,T}$. Expressions for the SM helicity amplitudes may be read off taking all $\alpha$'s or all $\beta$'s to zero. These SM results numerically match the output of \texttt{EvtGen}. 

Note that orthogonality of the $\DpiA$ and $\DpiB$ functions permit us to read off from the amplitudes which square and cross-terms contribute under integration over full angular phase space, and which are absent. For instance, the $f(q^2)\, g(q^2)$ cross-term integrates to zero. However, in the presence of angular phase space cuts, such terms do contribute. $D^*$ interference terms correspond to cross-terms within or between the $\DpiA$ or $\DpiB$ functions that contain orthogonal $\thD$ or $\phD$ dependence, and are typically $\mathcal{O}(1)$.

The decay $D^{*0} \to D^+\pi^-$ is kinematically forbidden, opening up a large $D^{*0} \to D^0\g$ branching ratio $\simeq 38\%$. This large branching ratio motivates consideration of the $B \to (D^{*} \to D \g)\tau \tn$ helicity amplitudes, too. We derive these amplitudes in Appendix~\ref{app:DDG}.

\subsection{\texorpdfstring{$\tau \to \ell \nu_\ell \tnbar$}{ttnlnl} and \texorpdfstring{$\tau \to \pi \tnbar$}{tptn}}
\label{sec:TDA}

Under the conventions of eq.~\eqref{eqn:TSBC}, the helicity amplitudes $[\mathcal{A}_{\tau \to \ell \nu_\ell \tnbar}]^{s_{\tn} }_{s_\tau}  \equiv [\mathcal{B}^{\ell}]^{s_{\tn}}_{s_\tau}$  for $\tau \to \ell \nu_\ell \tnbar$ are explicitly
\begin{subequations}
\begin{align}
	[\mathcal{B}^\ell]^\dn_1
		&  = i 2 \sqrt{2} G_F \sqrt{m_{\tau}^2 - p^2}\, \bigg\{m_{\tau} \cos\frac{\thW}{2} \sin\thell+2 e^{i(\phell - \phW)} \sqrt{p^2 } \cos^2\frac{\thell}{2} \sin\frac{\thW}{2}\bigg\}\,, \\[10pt]
	[\mathcal{B}^\ell]^\dn_2 
		 & = -i 2 \sqrt{2} e^{-i(\phtau - \phW)} G_F \sqrt{m_{\tau}^2 - p^2}\, \bigg\{2 e^{i(\phell - \phW)} \sqrt{p^2 } \cos^2\frac{\thell}{2} \cos\frac{\thW}{2}-m_{\tau} \sin\thell \sin\frac{\thW}{2}\bigg\}\,,
\end{align}
\end{subequations}
and $[\mathcal{B}^\ell]^\up_{1,2} = e^{i(\phtau - \phW)} [\mathcal{B}^\ell]^\dn_{1,2}$.  Note the quantum number, $s_{\tn}$, belonging to the $\tau$ neutrino in the parent $\bddstn$ process, is a consequence of our spinor phase conventions in eq.~\eqref{eqn:TSBC}, which ensures that $\phtau$ appears only in the physical combination $\phtau - \phW$.

For $\tau \to \pi \tnbar$, we adopt definitions for the helicity angles by replacing the $W$ with a pion in the $\tau$ decay within Fig.~\ref{fig:HASD}, and replacing $(\thW,\phW) \to (\thpi, \phpi) $ and $p^\mu \to p_\pi^\mu$. The helicity amplitudes $[\mathcal{A}_{\tau \to \pi \tnbar}]^{s_{\tn} }_{s_\tau}  \equiv [\mathcal{B}^\pi]^{s_{\tn}}_{s_\tau}$ are found to be
\begin{subequations}
\begin{align}
	[\mathcal{B}^\pi]^\dn_1 
		& = -i 2\sqrt{2} G_F f_\pi m_\tau\sqrt{m_\tau^2 - m_\pi^2} \cos \frac{\thpi}{2}\,,\\[10pt]
	[\mathcal{B}^\pi]^\dn_2 
		& = -i 2 \sqrt{2} G_F f_\pi m_\tau e^{-i(\phtau - \phpi)}\sqrt{m_\tau^2 - m_\pi^2} \sin \frac{\thpi}{2}\,,
\end{align}
\end{subequations}
and $[\mathcal{B}^\pi]^\up_{1,2} = e^{i(\phtau - \phpi)} [\mathcal{B}^\pi]^\dn_{1,2}$. Here $f_\pi = 93$\,MeV is the pion decay constant.

\section{Applications}
\label{sec:apps}

The computation of the NP helicity amplitudes for $B \to \dds( \to DY) \tau (\to X \tnbar) \tn$ decays permits us to efficiently reweigh large Monte Carlo samples to any theory generated by the NP operators~\eqref{eqn:FFOD}. We may thereby access the full kinematic structure of the (visible) $\tau$ and $D^*$ decay products, and explore the NP effects therein. To illustrate the potential usefulness and NP discrimination power of these results, in this section we provide a first exploration of such NP effects for $B \to D^*(\to D \pi)\tau(\to \ell \nu_\ell \tnbar)\tn$, focusing on NP scenarios compatible with the $\bddstn$ rate~\cite{Freytsis:2015qca}. We include effects of $q^2$, missing momentum, and lepton energy cuts in this analysis. However, background modelling, detector simulations, or $B \to D \tau \tn$ pollution, all of which are required for a realistic analysis, are deferred to a future study~\cite{Hammer:2016}. 

\subsection{Monte Carlo strategy}

In accordance with the results of Sec.~\ref{sec:HA}, the full $B \to D^*( \to D\pi)\tau(\to \ell \nu_\ell \tnbar) \tn$ helicity amplitudes may be expressed in the linear form 
\begin{equation}
	[\mathcal{M}]^{s_{\tn}}_{s_{\tnbar} s_\ell s_{\nu_\ell}} = \vec{v}\cdot [\vec{\mathcal{M}}_v]^{s_{\tn}}_{s_{\tnbar} s_\ell s_{\nu_\ell}}\,, 
\end{equation}
where $\vec{\mathcal{M}}_v$ is a \emph{vector of amplitudes} and the 11-dimensional vector $\vec{v}$ is
\begin{multline}
	\vec{v} = \Big( 1\,,\ \alSR\beSL r_S^2\,,\ \alSR\beSR r_S^2\,,\ \alSL\beSL r_S^2\,,\ \alSL\beSR r_S^2\,,\ \alVR\beVL r_V^2\,,\ \\
		 \alVR\beVR r_V^2\,,\ \alVL\beVL r_V^2\,,\ \alVL\beVR r_V^2\,,\ \alTR\beTL r_T^2\,,\ \alTL\beTR r_T^2 \Big)\,.
\end{multline}
The first entry of $\vec{\mathcal{M}}_v$ corresponds to the SM contribution. By construction, $\vec{\mathcal{M}}_v$ is independent of the particular NP model, but depends only on phase space configuration. Our MC strategy is then as follows: (i) A large MC sample of pure phase space weighted events is created; (ii) For each event, the Hermitian matrix of weights $W_v \equiv \vec{\mathcal{M}}_v(\vec{\mathcal{M}}_v)^\dagger $ is computed from the results in Sec.~\ref{sec:HA}; (iii) These matrix weights are then either 1D, 2D or $n$D histogrammed with respect to a set of kinematic observables $\mathcal{O}_i$, or alternatively, the matrix weights are collated event-by-event with the observables $\mathcal{O}_i$; (iv) After all reweighting, the histograms or weighted event sample corresponding to a particular NP point may be generated by contracting all matrix weights with the desired $\vec{v}$, i.e., via $\vec{v}\,^\dagger W_v \vec{v}$. 

At present, step (i) is performed with \texttt{EvtGen}~\cite{Lange:2001uf}, while steps (ii) and (iii) are executed by our own \texttt{Python} code. In this strategy, reweighting of the MC sample into matrix weights, $W_v$, need be performed only once for any given choice of phase space cuts, while ranging over the multi-dimensional space of NP couplings is reduced to the highly efficient post-reweighting linear operation, $\vec{v}\,^\dagger W_v \vec{v}$. We therefore just use \texttt{Mathematica} for step (iv). The amplitude-level calculation of $\vec{\mathcal{M}}_v$ permits calculation of the $11\times11$ weight matrix, $W_v$, with roughly an order of magnitude fewer floating point operations than a direct amplitude-squared calculation, and therefore makes practical the reweighting of large MC samples for multiple cut choices. 

We shall consider here an MC sample of $10$~million events, reweighted once on the full phase space, and once with application of the phase space cuts, motivated by Refs.~\cite{Lees:2013uzd,Huschle:2015rga},
\begin{equation}
	\label{eqn:PSC}
	E_{\ell} > 400~\text{MeV}\,,\qquad m^2_{\text{miss}} > 1.5~\text{GeV}^2\,,\qquad q^2 > 4~\text{GeV}^2\,.
\end{equation}
With three neutrinos in the final state, the remaining visible phase space for $B \to D^*(\to D \pi)\tau(\to \ell \nu_\ell \tnbar)\tn$ is parametrized by seven independent parameters. In the $B$ rest frame we compute an overcomplete set of ten observables, including
\begin{equation}
	\label{eqn:LKO}
	q^2\,, \quad E_{D^*}\,, \quad E_D\,, \quad E_{\pi}\,, \quad E_{\ell}\,, \quad \cos\theta_{D\pi}, \quad \cos\theta_{\pi\ell}, \quad \cos\theta_{D\ell}\,,
\end{equation}
where $\cos \theta_{XY}$ is the opening angle between $\bm{p}_X$ and $\bm{p}_Y$, as well as the normalized triple product and the missing invariant mass, respectively,
\begin{equation}
	\label{eqn:LKO2}
	V_{D\pi\ell} \equiv \hat{\bm{p}}_D \cdot (\hat{\bm{p}}_{\pi} \times \hat{\bm{p}}_{\ell})\,, \qquad \text{and} \qquad m^2_{\rm miss} \equiv (k_{\tn} + k_{\tnbar} + k_{\nu_\ell})^2\,.
\end{equation}
To generate the $\bdstn$ form factors~\eqref{eqn:FFDS}, we use the ISGW2 parametrization~\cite{Scora:1995ty,Isgur:1988gb} for $f(q^2)$ as presently implemented in \texttt{EvtGen}~\cite{Lange:2001uf,Ryd:2005zz} and obtain the $q^2$-dependence of the rest via the leading order HQET relations~\eqref{eqn:HQETDS}.

\subsection{Univariate versus bivariate analyses}

Various NP scenarios may produce $\bddstn$ rates commensurate with the central values of current observations. In particular, leptoquark models with couplings
\begin{equation}
	\alTR\beTL r_T^2 = -0.38\,,\qquad \alTR\beTL r_T^2 = 0.05\,,\qquad \big\{\alTR\beTL r_T^2\,,~ \alSR\beSL r_S^2 \big\}  = \{-0.04\,,~0.16\}\,,
\end{equation}
can reproduce the central values of the observed $\bddstn$ rates~\cite{Freytsis:2015qca}. (In the notation of Ref.~\cite{Freytsis:2015qca}, these values correspond to the Wilson coefficients $C_T = 0.52(\Lambda/1~\text{TeV})^2$, $C_T= -0.07(\Lambda/1~\text{TeV})^2$ and $C_{S_L}'' = -0.46(\Lambda/1~\text{TeV})^2$, respectively.)

\begin{figure}[p]
\centering{\hfill
\includegraphics[width = 0.4\linewidth]{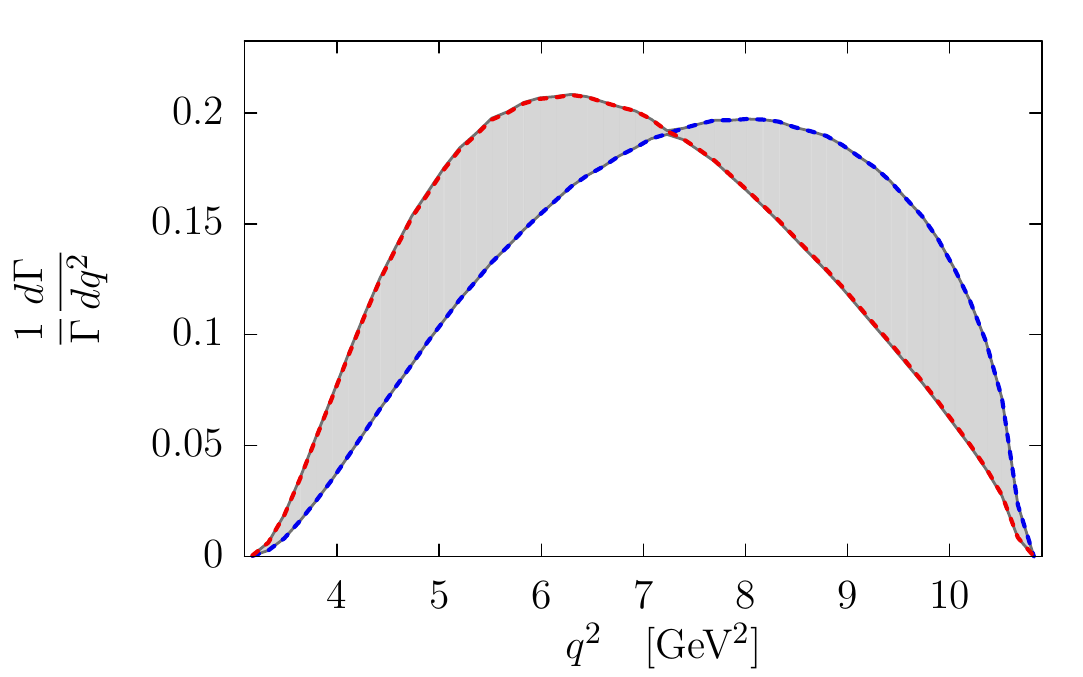}\hfill
\includegraphics[width = 0.4\linewidth]{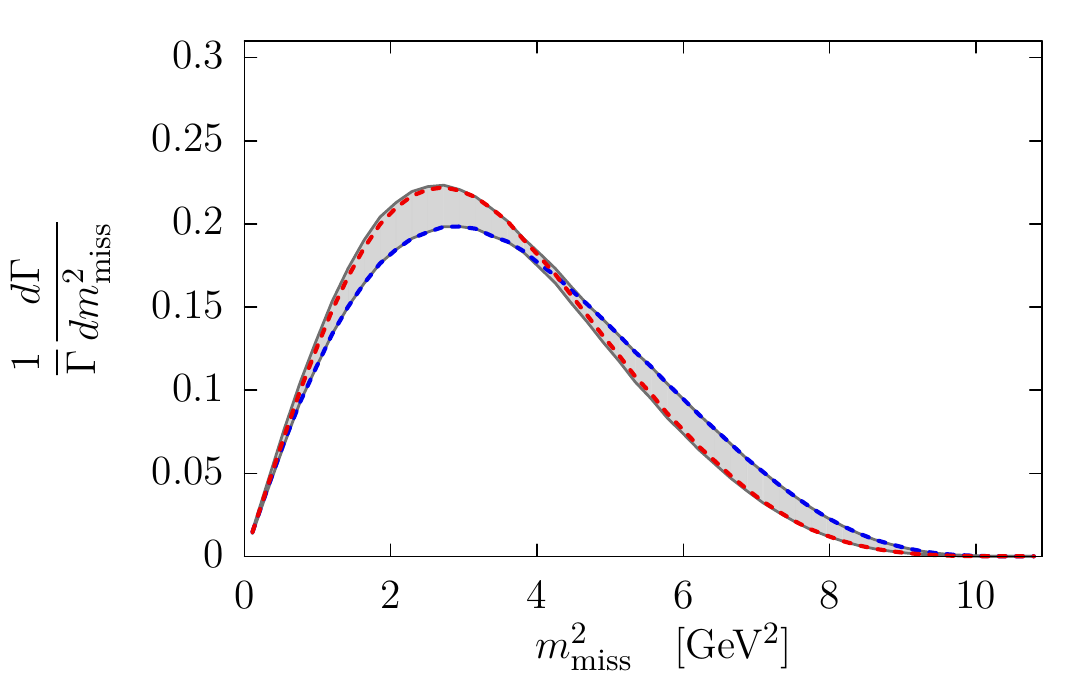}  \hfill \\ 
\hfill
\includegraphics[width = 0.4\linewidth]{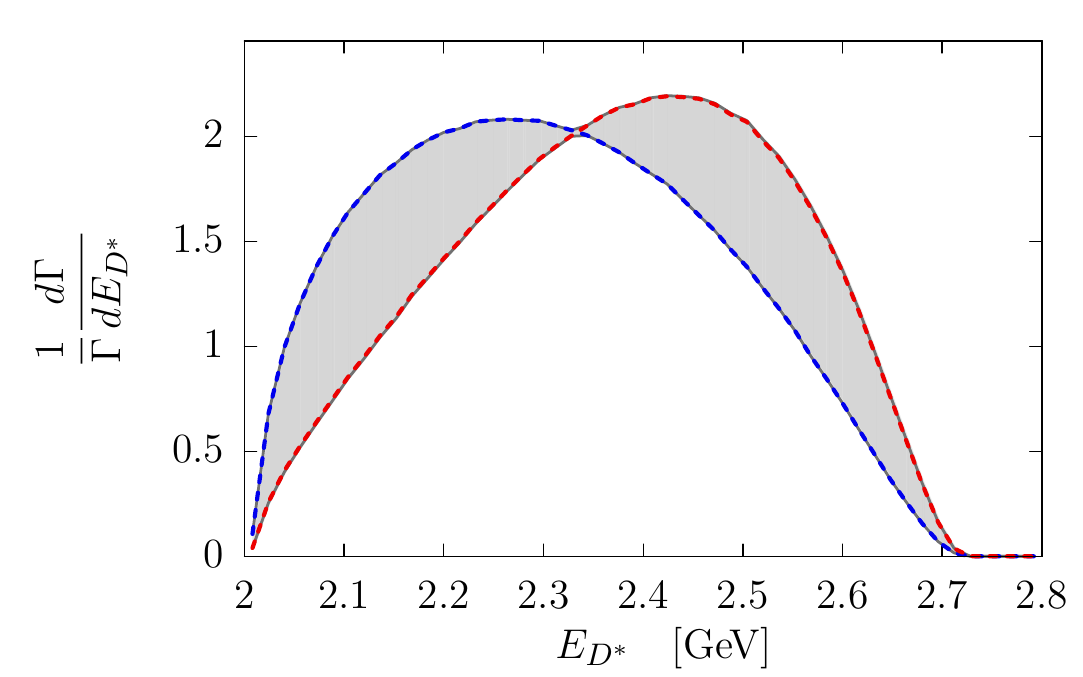} \hfill
\includegraphics[width = 0.4\linewidth]{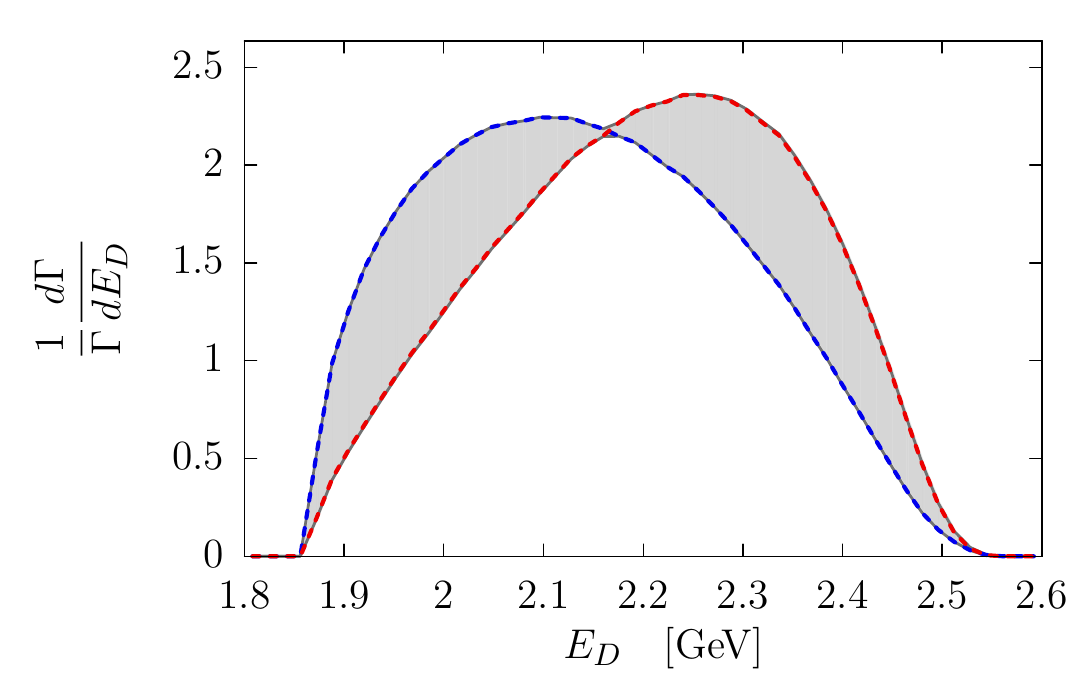} \hfill \\ 
\hfill
\includegraphics[width = 0.4\linewidth]{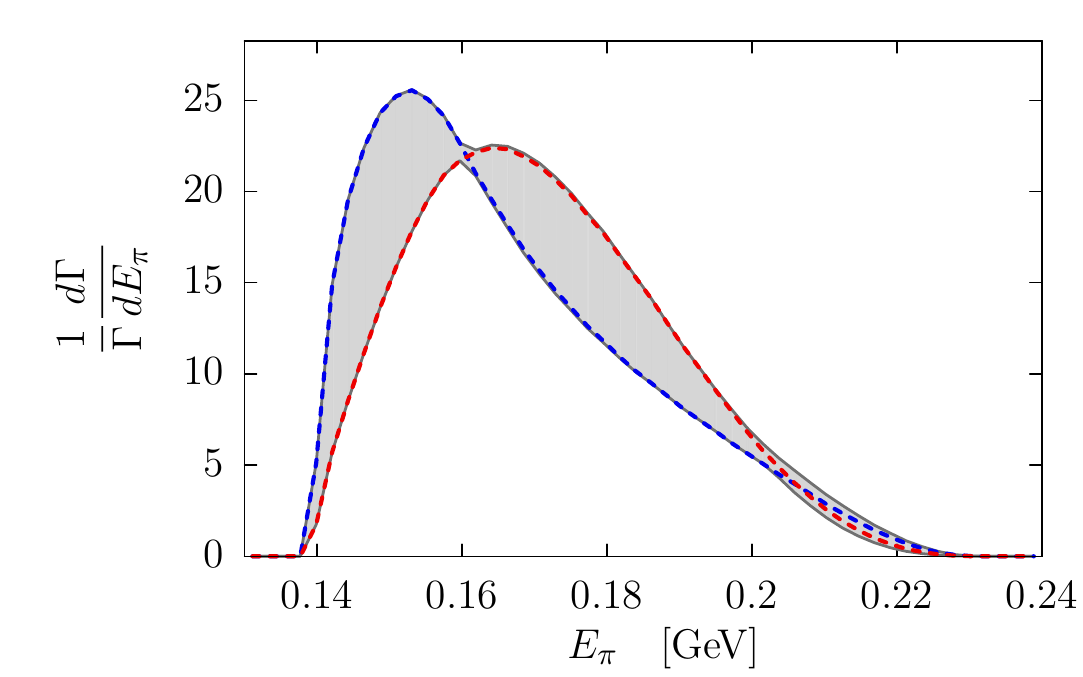} \hfill
\includegraphics[width = 0.4\linewidth]{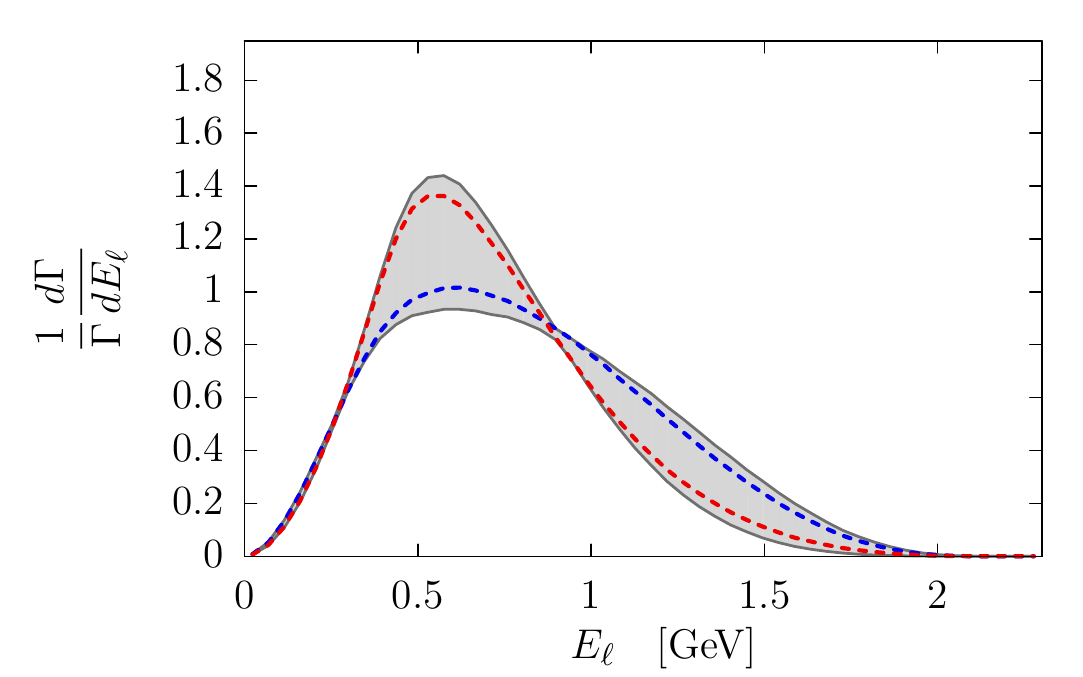} \hfill \\ 
\hfill
\includegraphics[width = 0.4\linewidth]{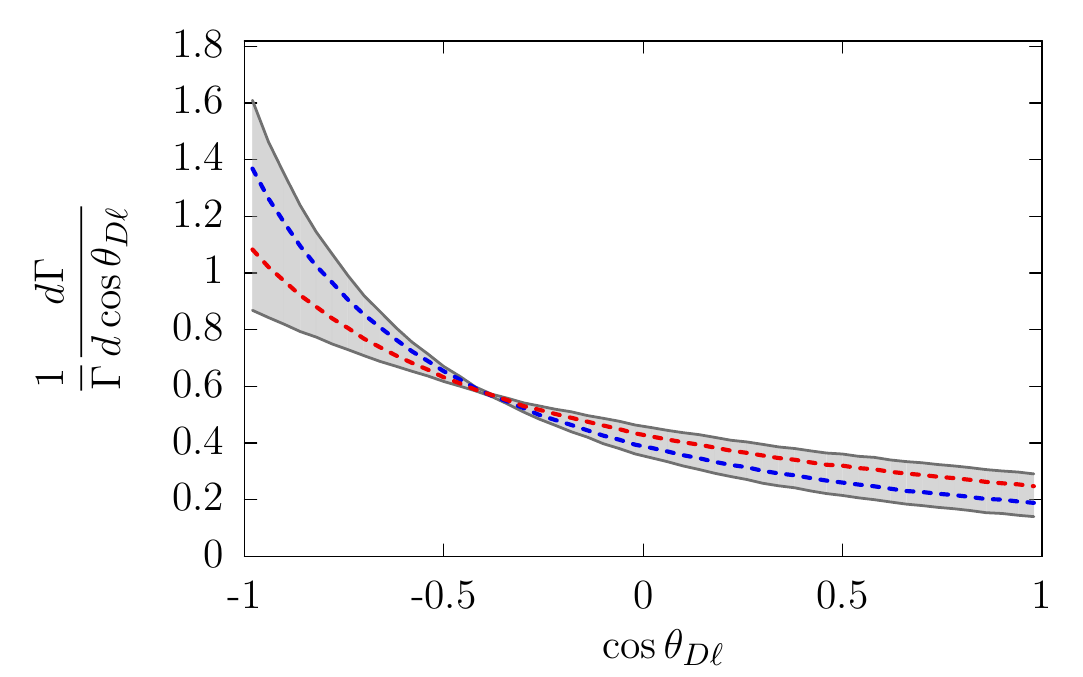} \hfill
\includegraphics[width = 0.4\linewidth]{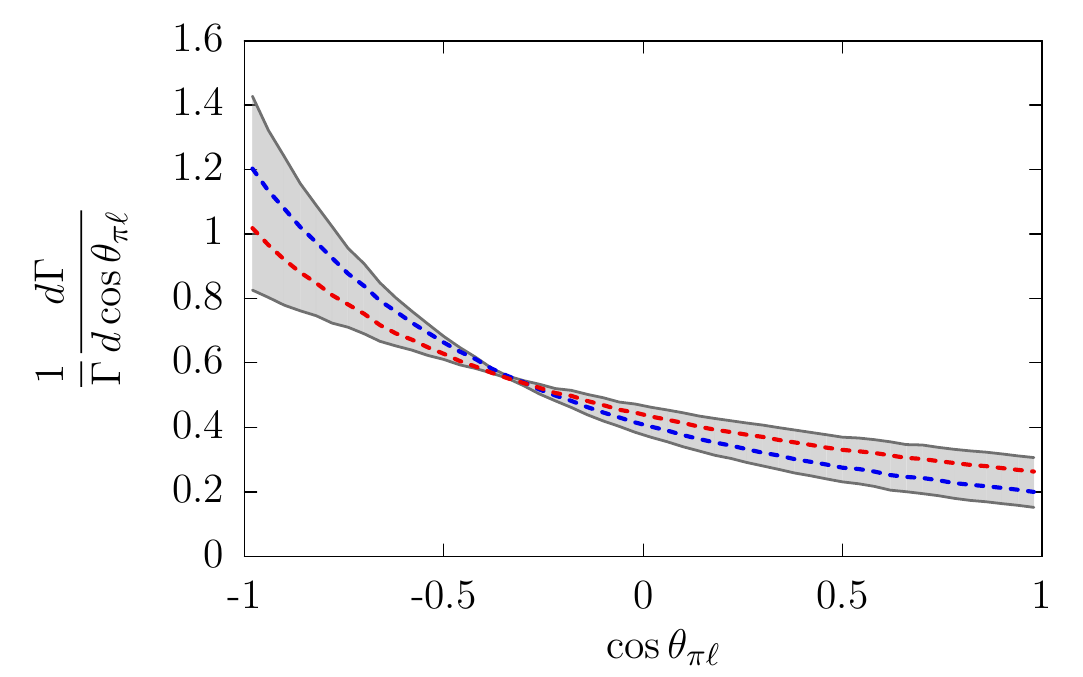} \hfill \\ 
\hfill
\includegraphics[width = 0.4\linewidth]{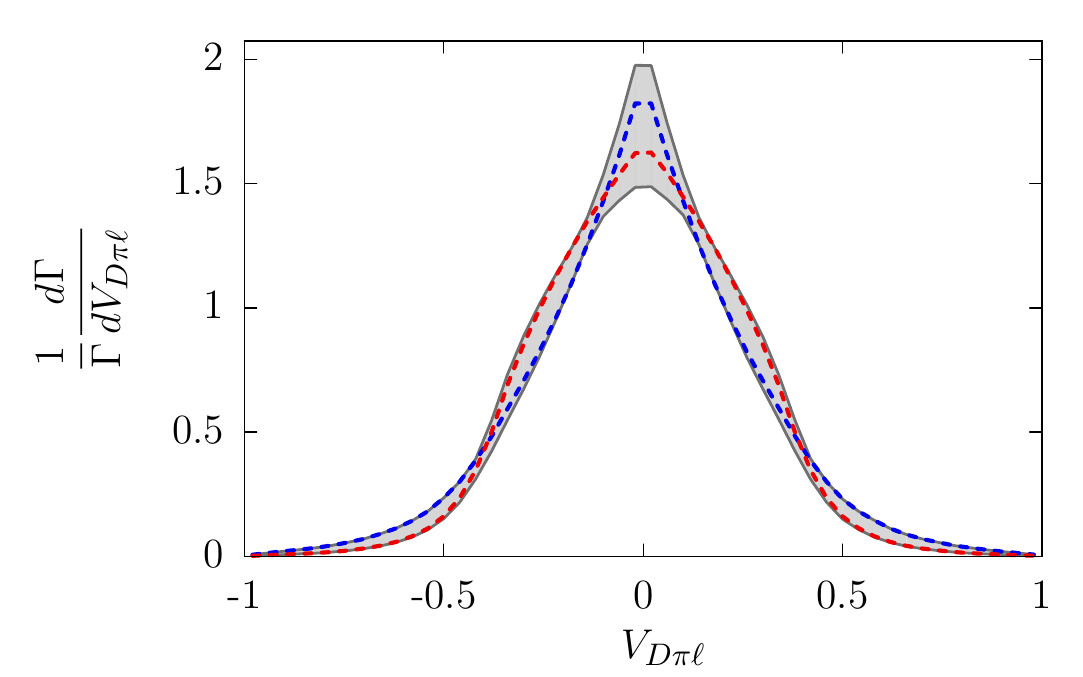} \hfill
\includegraphics[width = 0.4\linewidth]{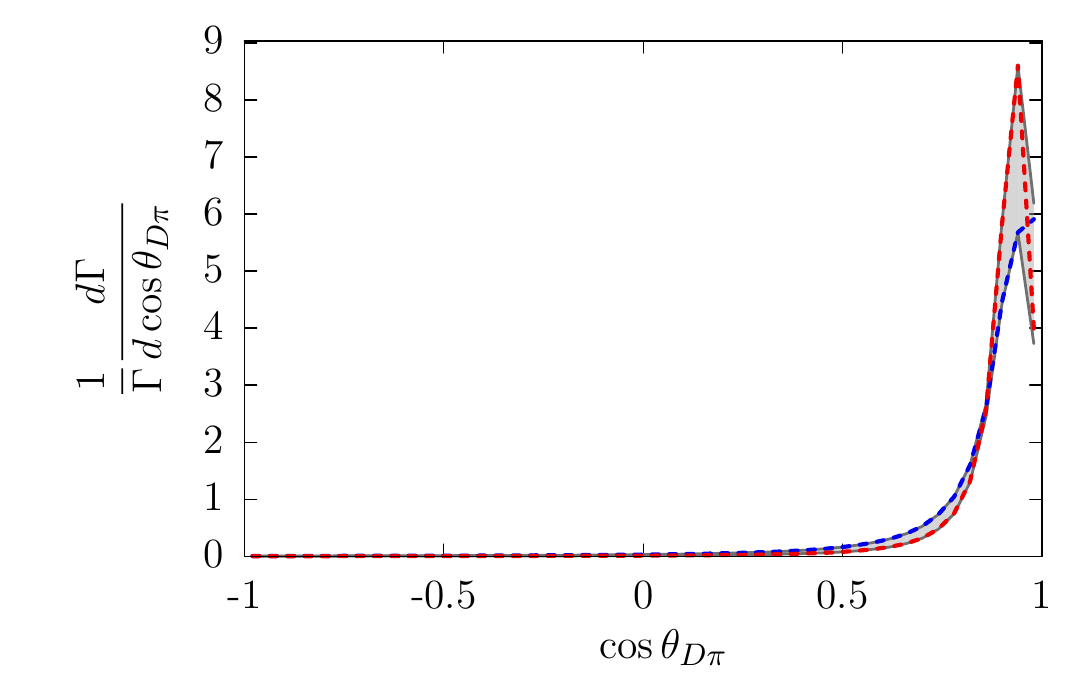} 
}
\caption{Kinematic distributions in the $B$ rest frame for couplings ranging over $g_T \in [-0.76, 0]$ (gray regions) without phase space cuts. The blue (red) dashed curves show the SM ($g_T = -0.38$).}
\label{fig:1DH}
\end{figure}

\begin{figure}[p]
\centering{\hfill
\includegraphics[width = 0.4\linewidth]{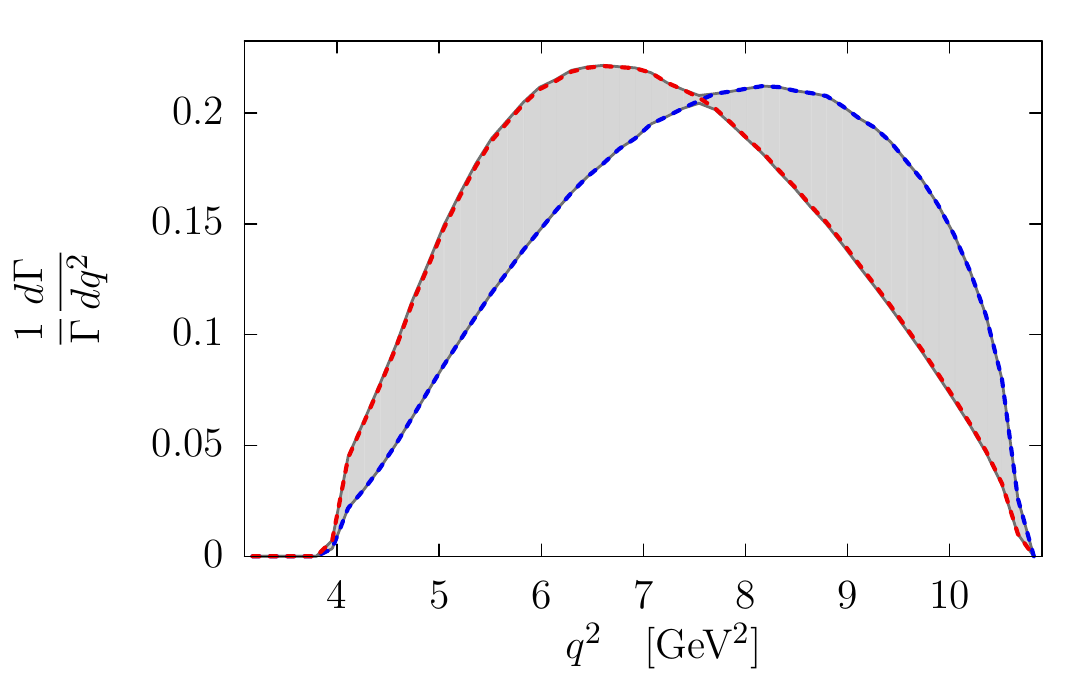}\hfill
\includegraphics[width = 0.4\linewidth]{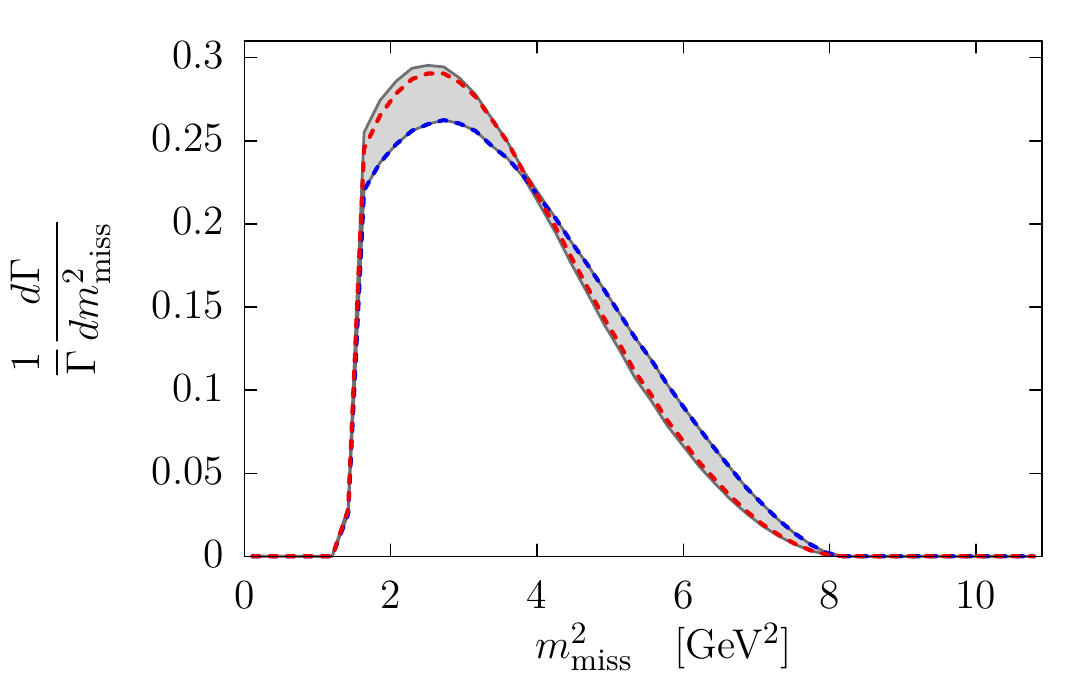} \hfill \\ 
\hfill
\includegraphics[width = 0.4\linewidth]{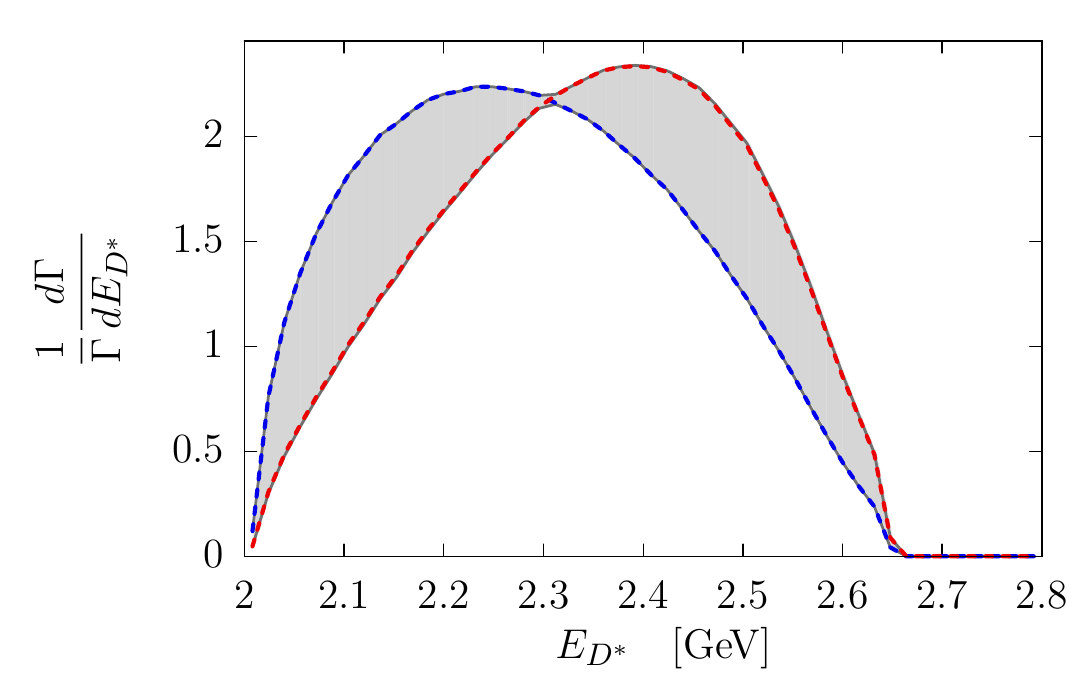} \hfill
\includegraphics[width = 0.4\linewidth]{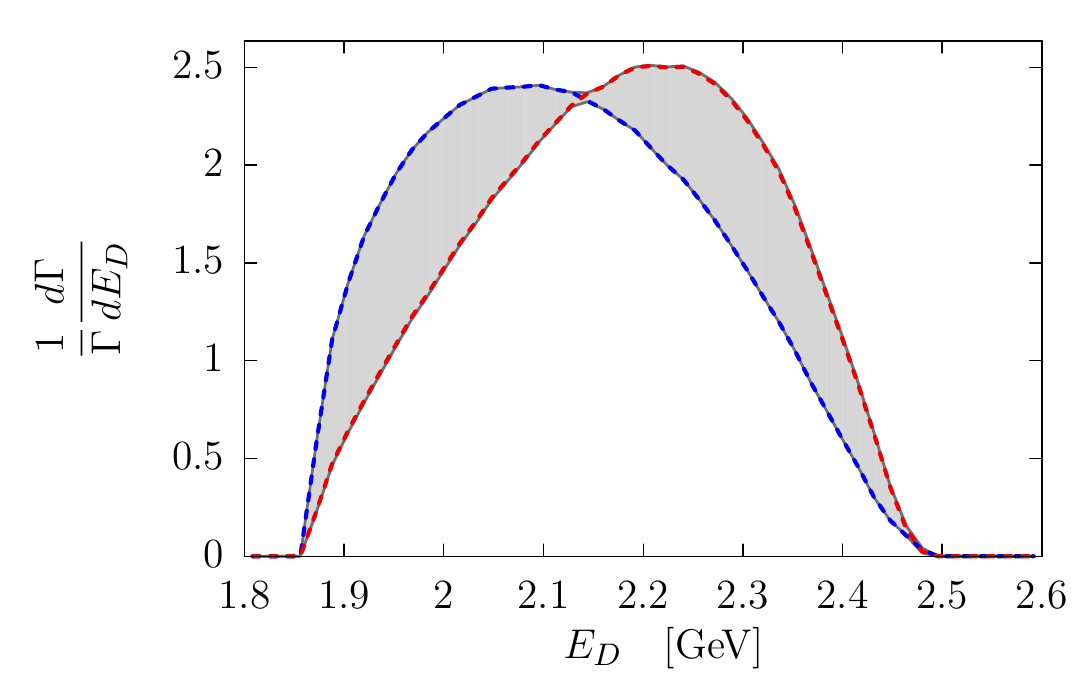} \hfill \\ 
\hfill
\includegraphics[width = 0.4\linewidth]{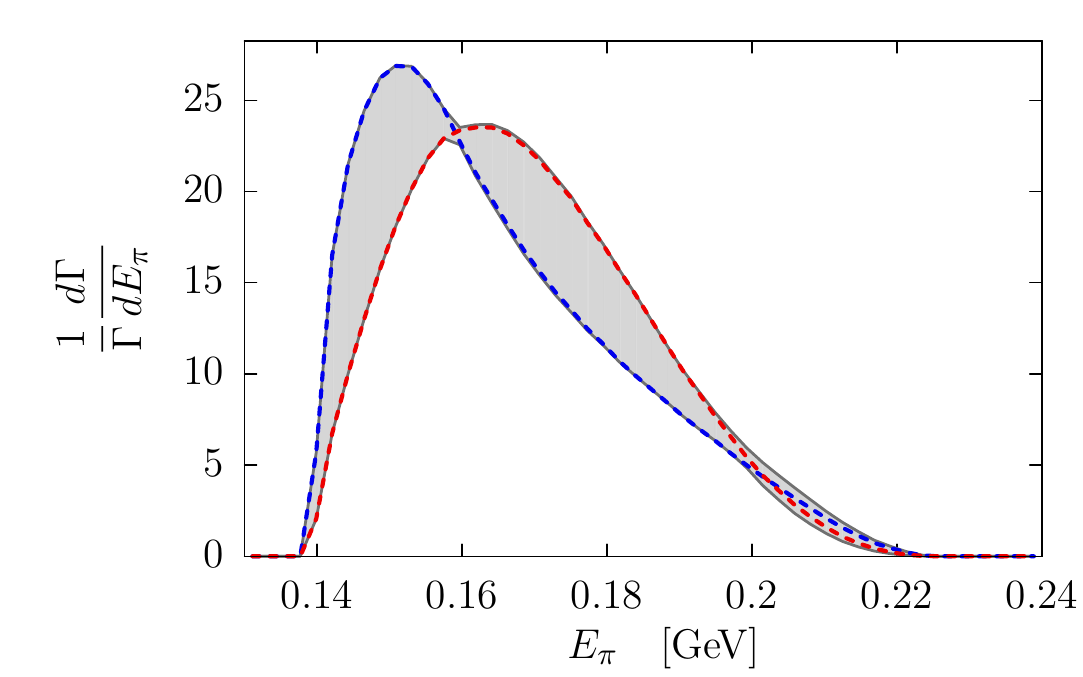} \hfill
\includegraphics[width = 0.4\linewidth]{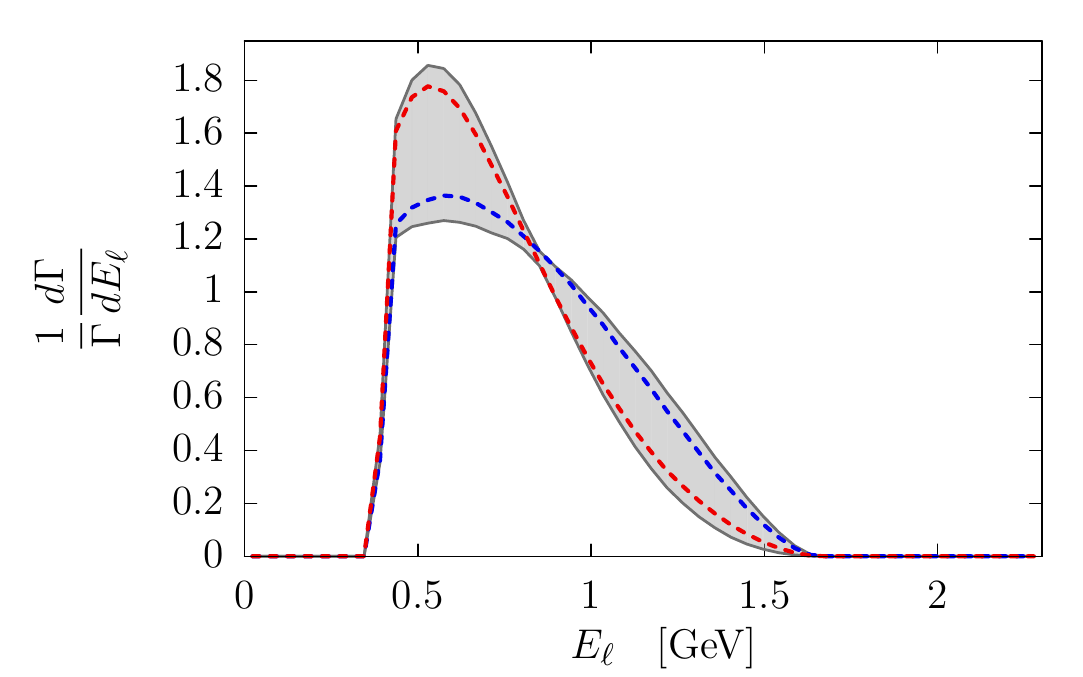} \hfill \\ 
\hfill
\includegraphics[width = 0.4\linewidth]{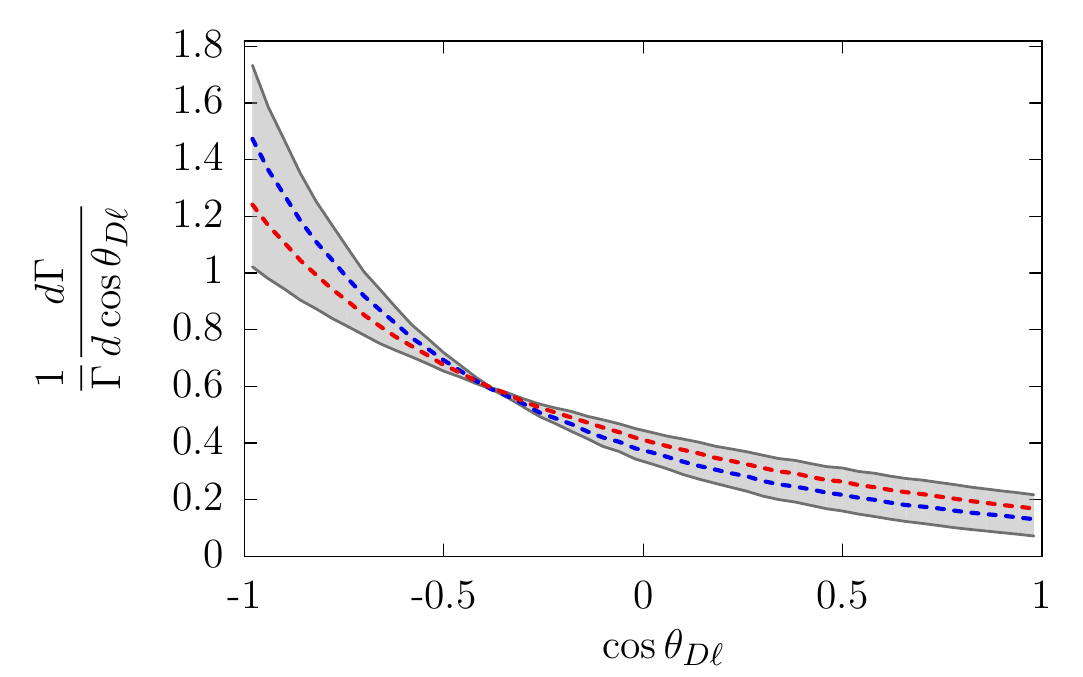} \hfill
\includegraphics[width = 0.4\linewidth]{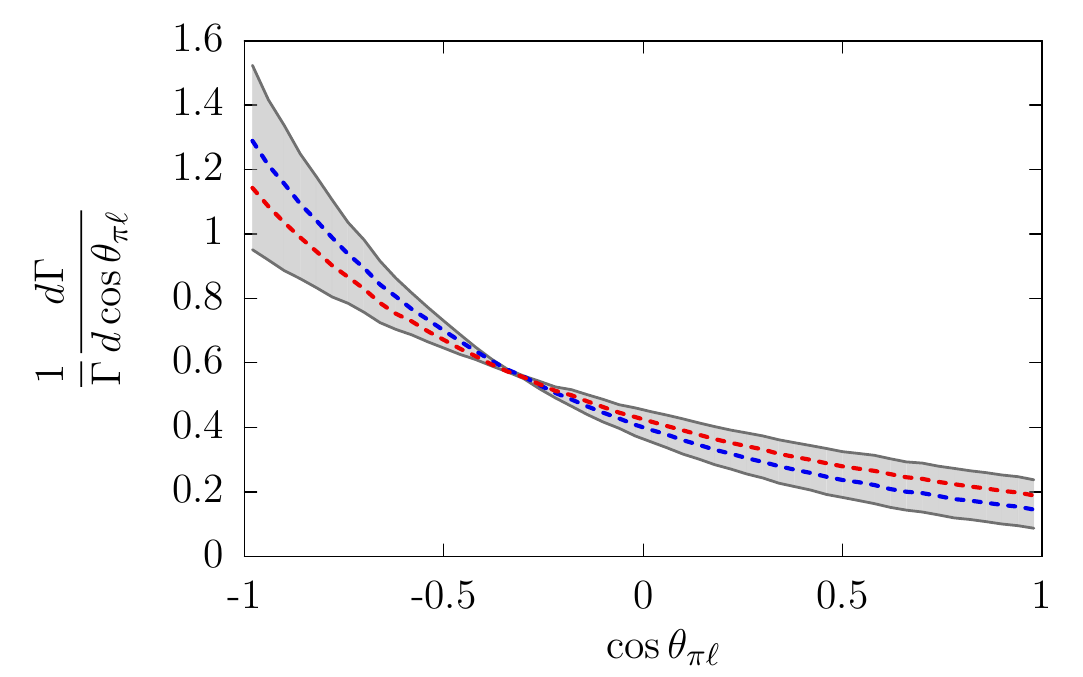} \hfill \\ 
\hfill
\includegraphics[width = 0.4\linewidth]{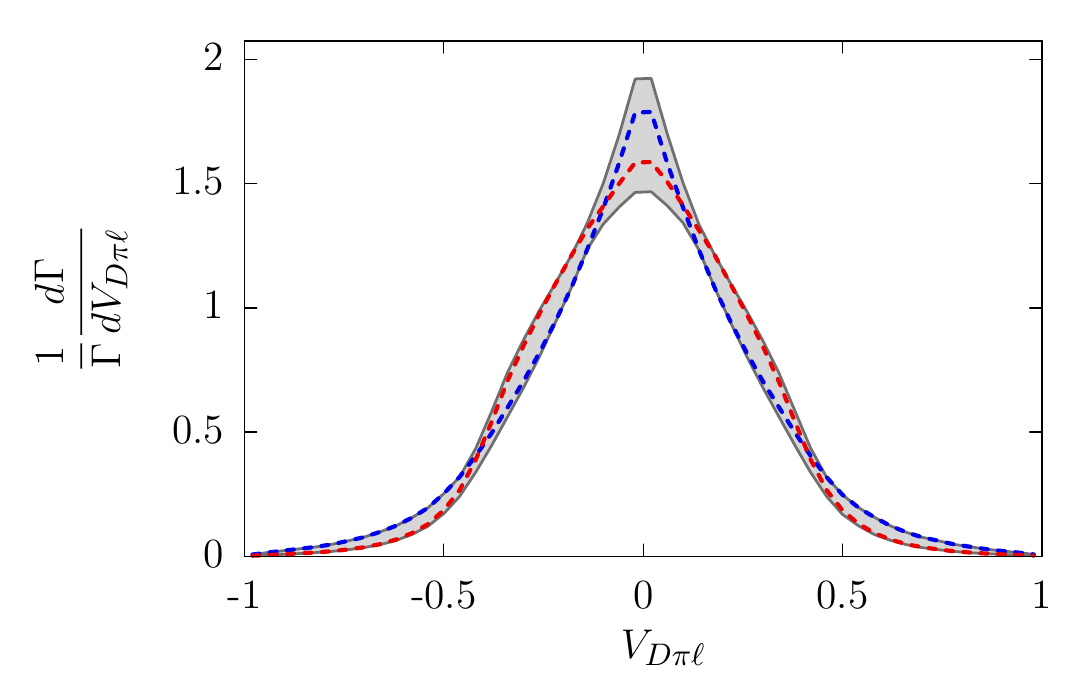} \hfill
\includegraphics[width = 0.4\linewidth]{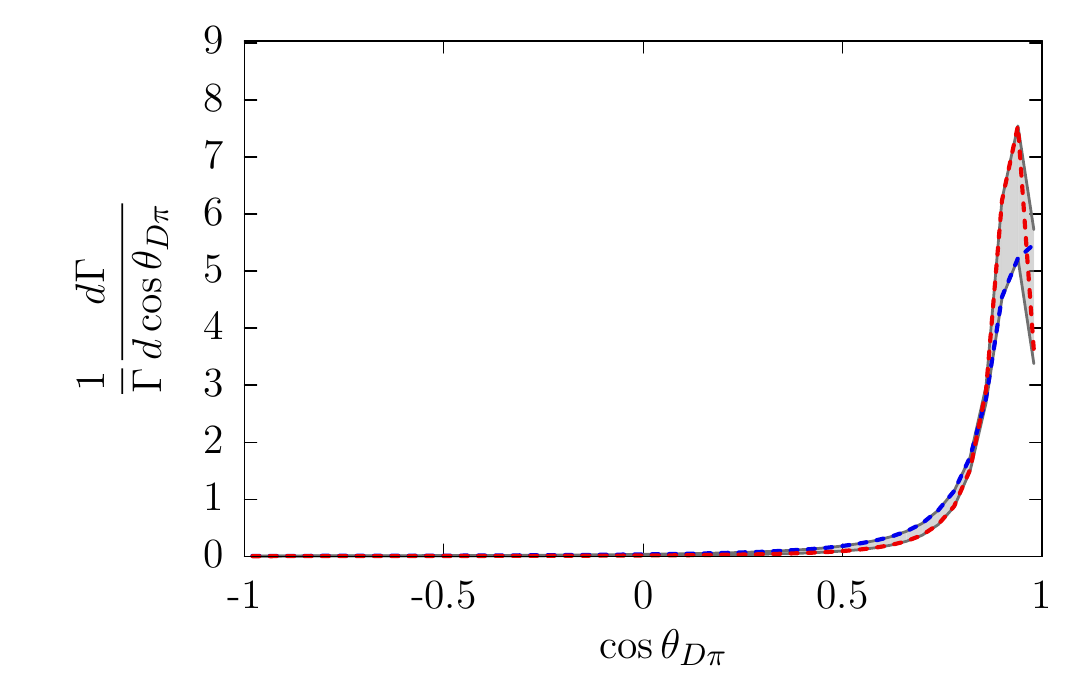}
}
\caption{Kinematic distributions in the $B$ rest frame for couplings ranging over $g_T \in [-0.76, 0]$ (gray regions) with phase space cuts~\eqref{eqn:PSC}. The blue (red) dashed curves show the SM ($g_T = -0.38$).}
\label{fig:1DHC}
\end{figure}

In this section, as an example, we focus on the NP model with $g_T \equiv \alTR\beTL r_T^2 = -0.38$.  In Figs.~\ref{fig:1DH} and \ref{fig:1DHC}, we present the differential distributions for each of the ten kinematic observables~\eqref{eqn:LKO}--\eqref{eqn:LKO2} in the full and cut phase space, respectively, generated by ranging over $g_T \in [-0.76, 0]$, i.e., over a range spanning twice the best fit $g_T$ value. We also show the distributions for $g_T = -0.38$ and the SM. While the $q^2$ distribution itself has some discriminating power between the SM and the NP along the $g_T$ contour, other observables, in particular $E_\ell$, $\cos\theta_{D\ell}$, and $\cos\theta_{\pi\ell}$ may be just as, if not more, discriminating.

\begin{figure}[t]
\includegraphics[width = 0.32\linewidth]{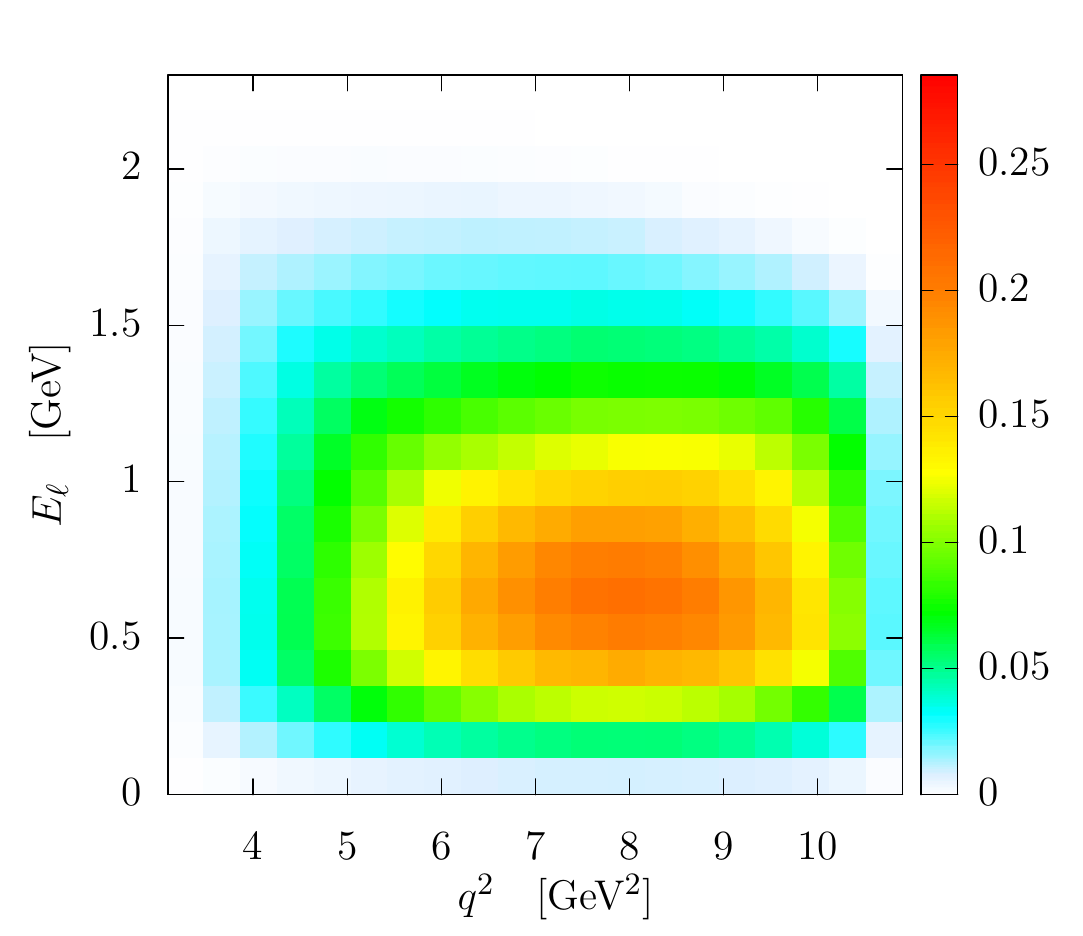}\hfill
\includegraphics[width = 0.32\linewidth]{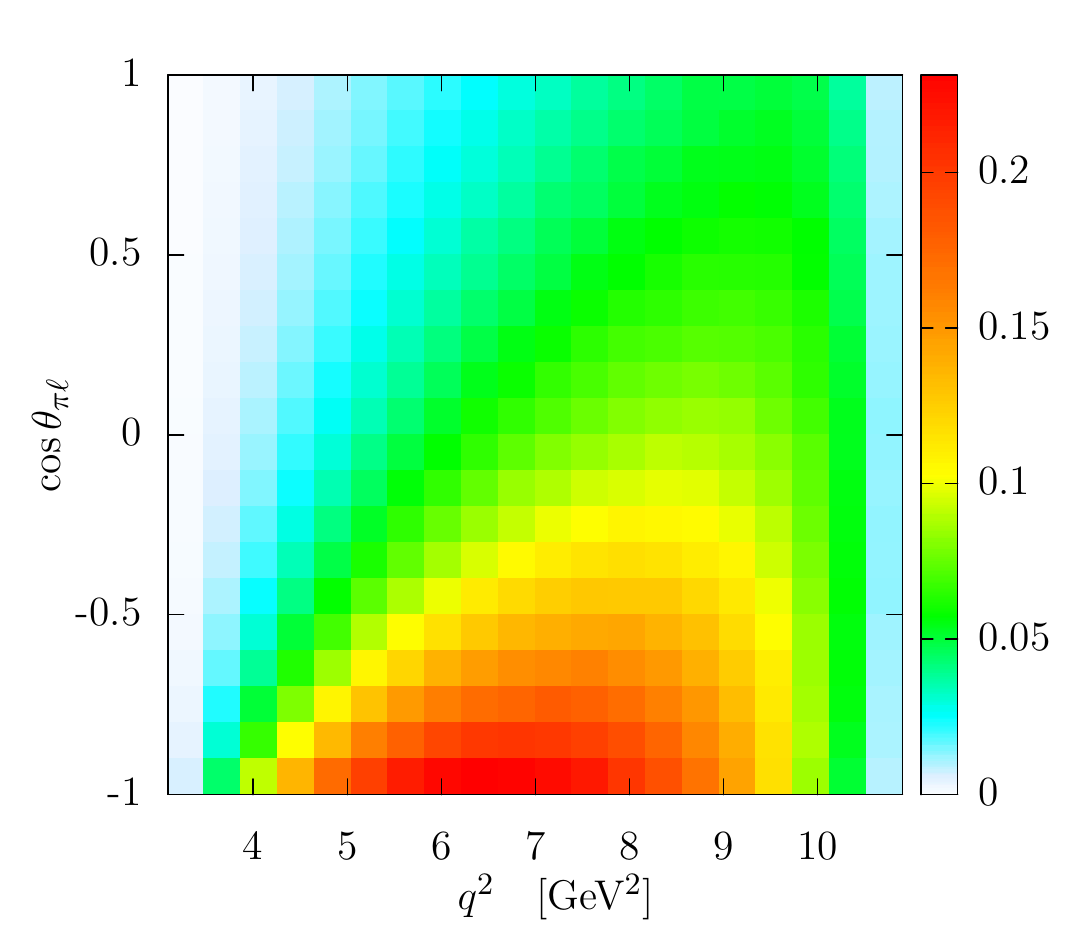}\hfill
\includegraphics[width = 0.32\linewidth]{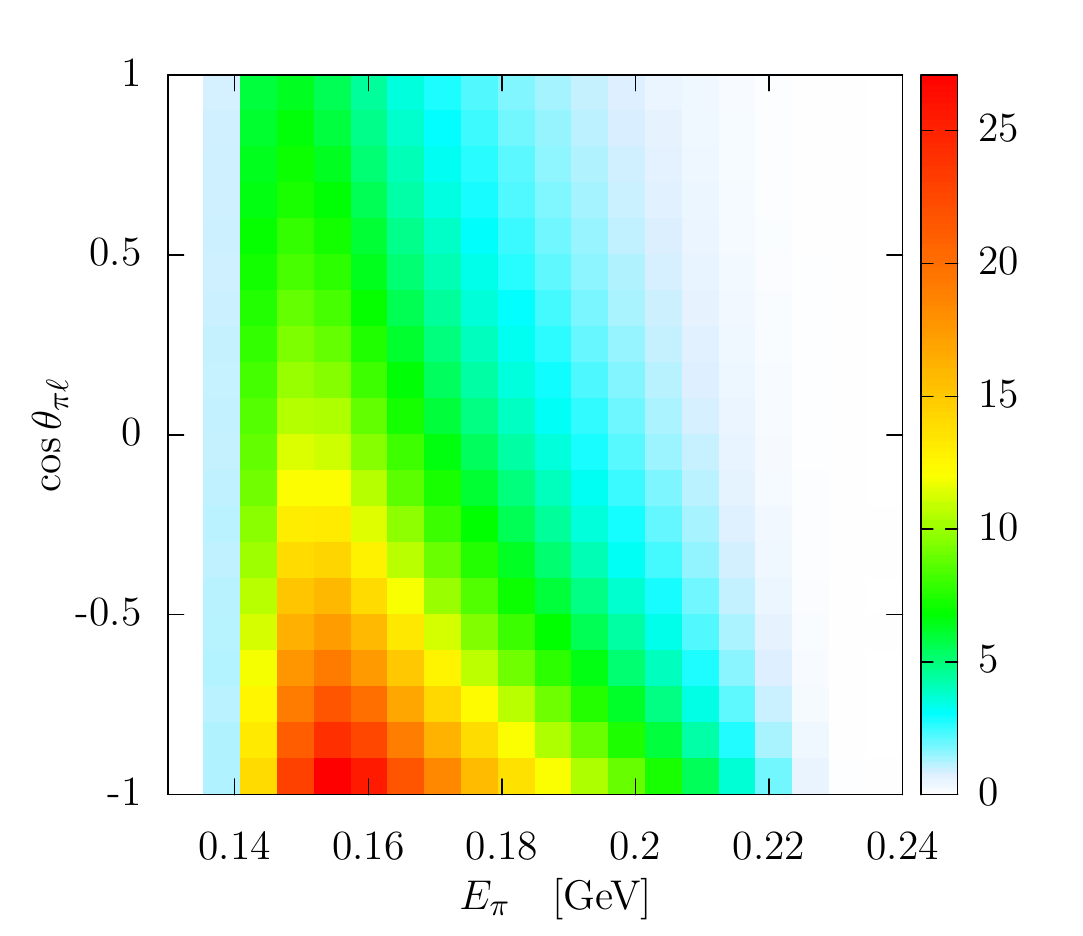} \\
\includegraphics[width = 0.32\linewidth]{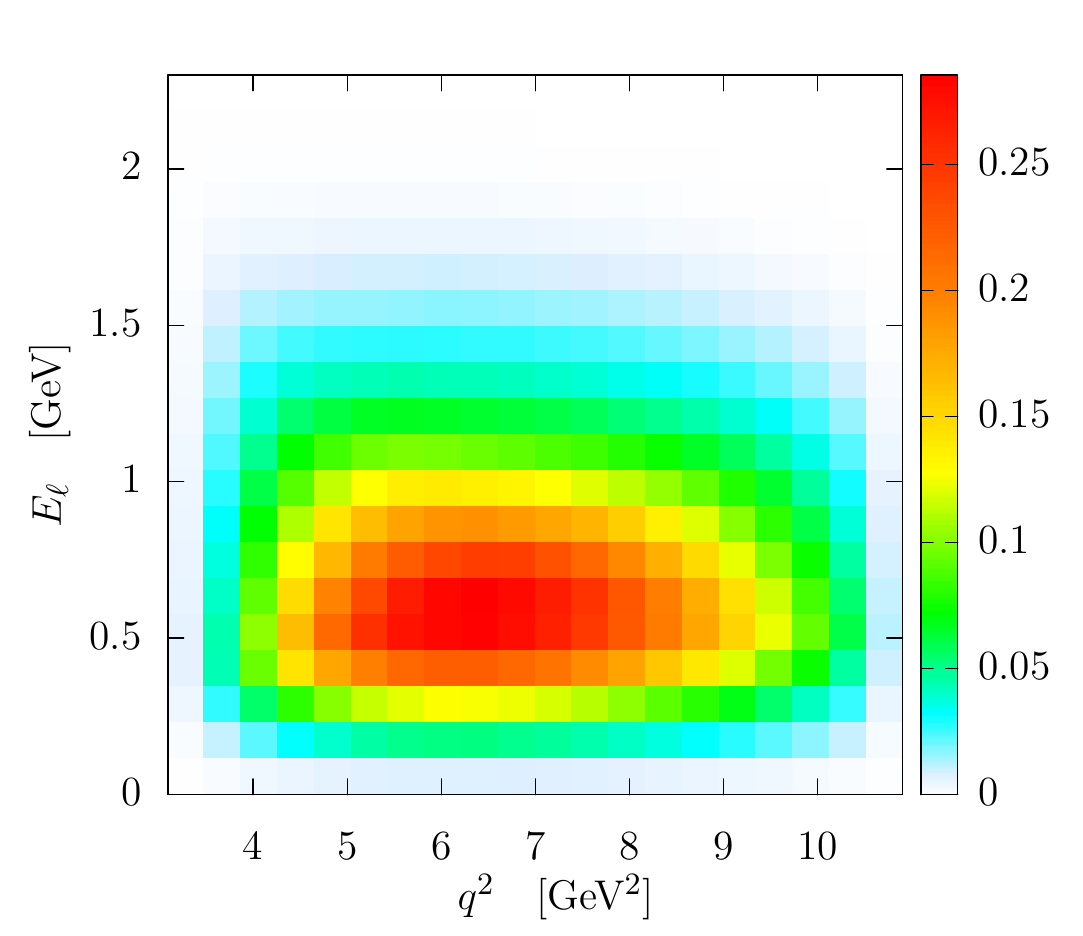}\hfill
\includegraphics[width = 0.32\linewidth]{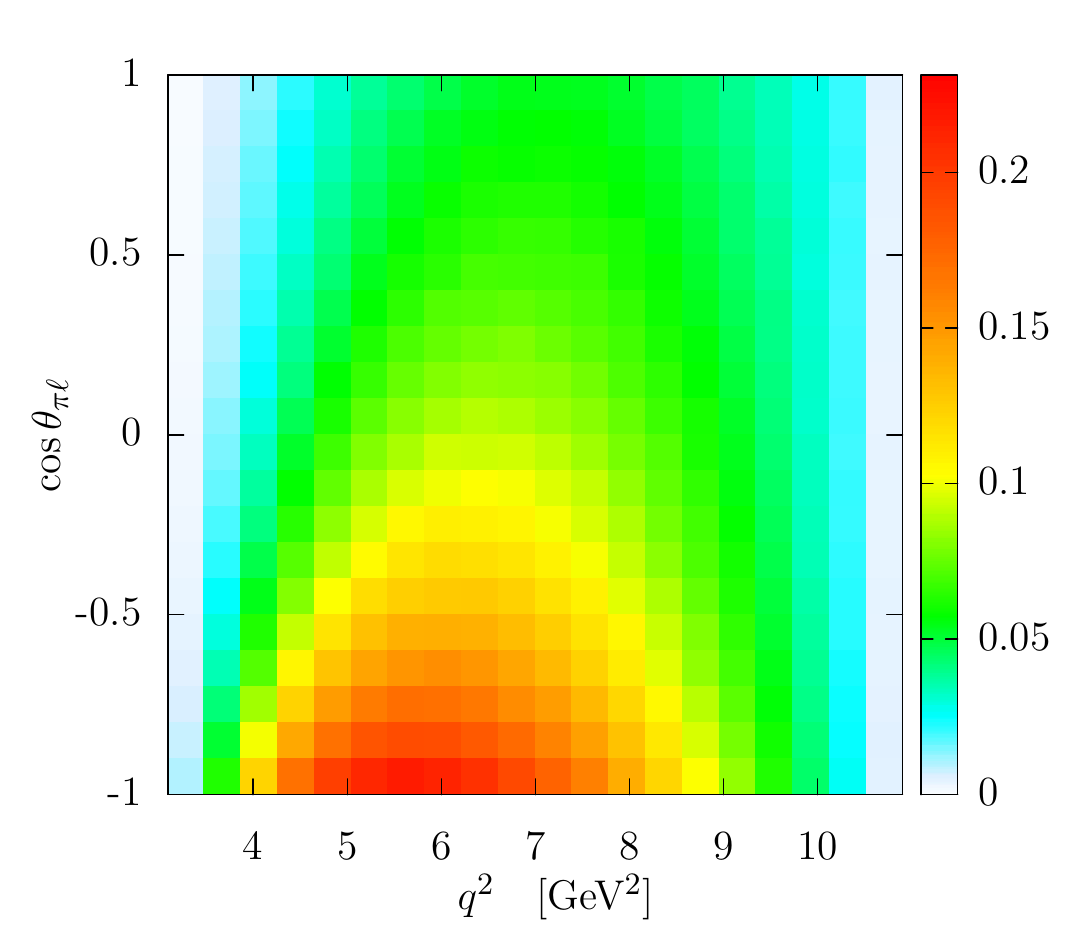}\hfill
\includegraphics[width = 0.32\linewidth]{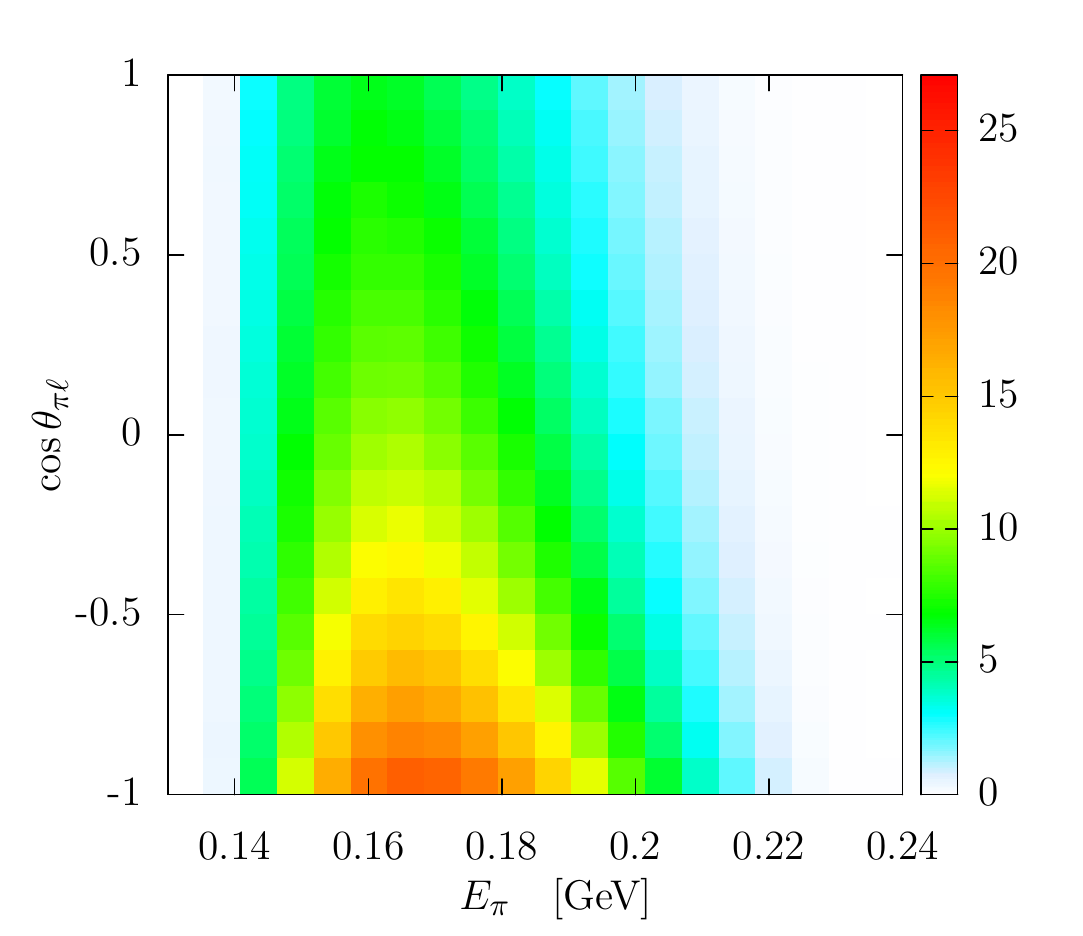} \\
\includegraphics[width = 0.32\linewidth]{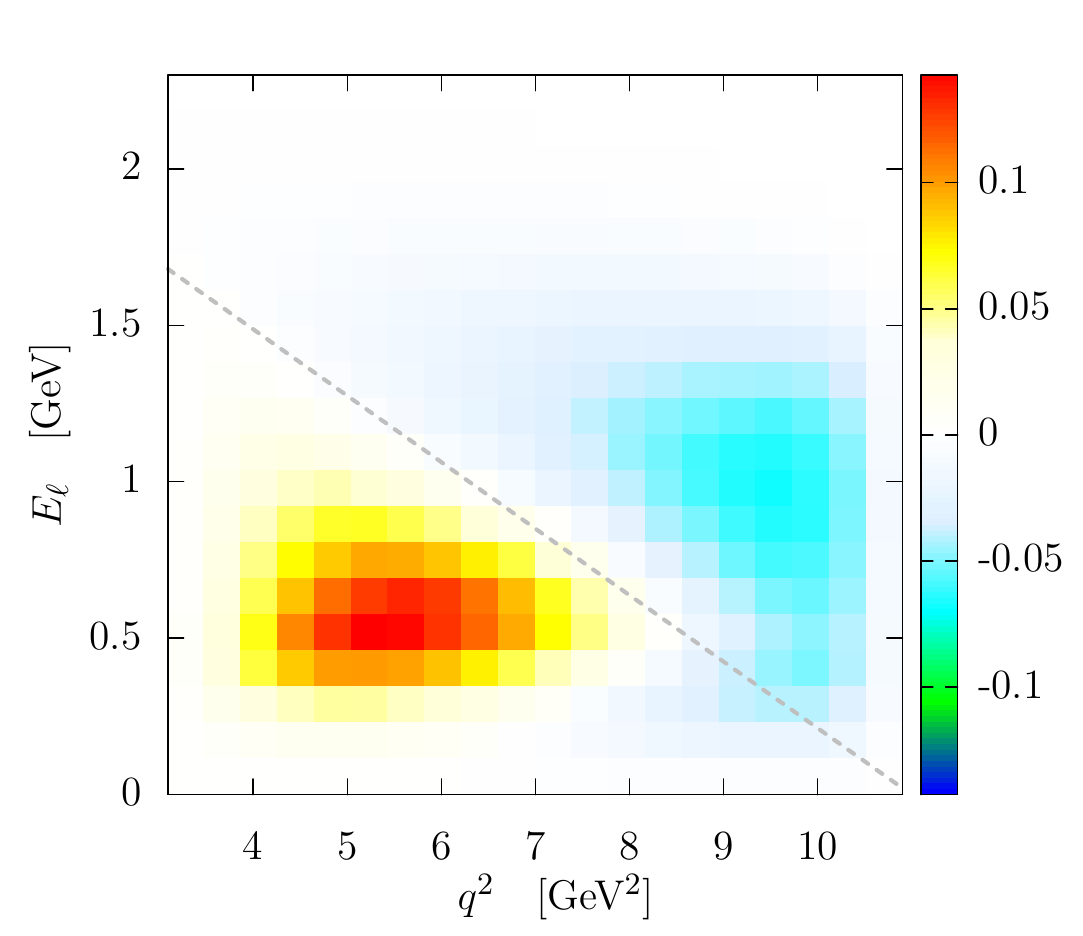}\hfill
\includegraphics[width = 0.32\linewidth]{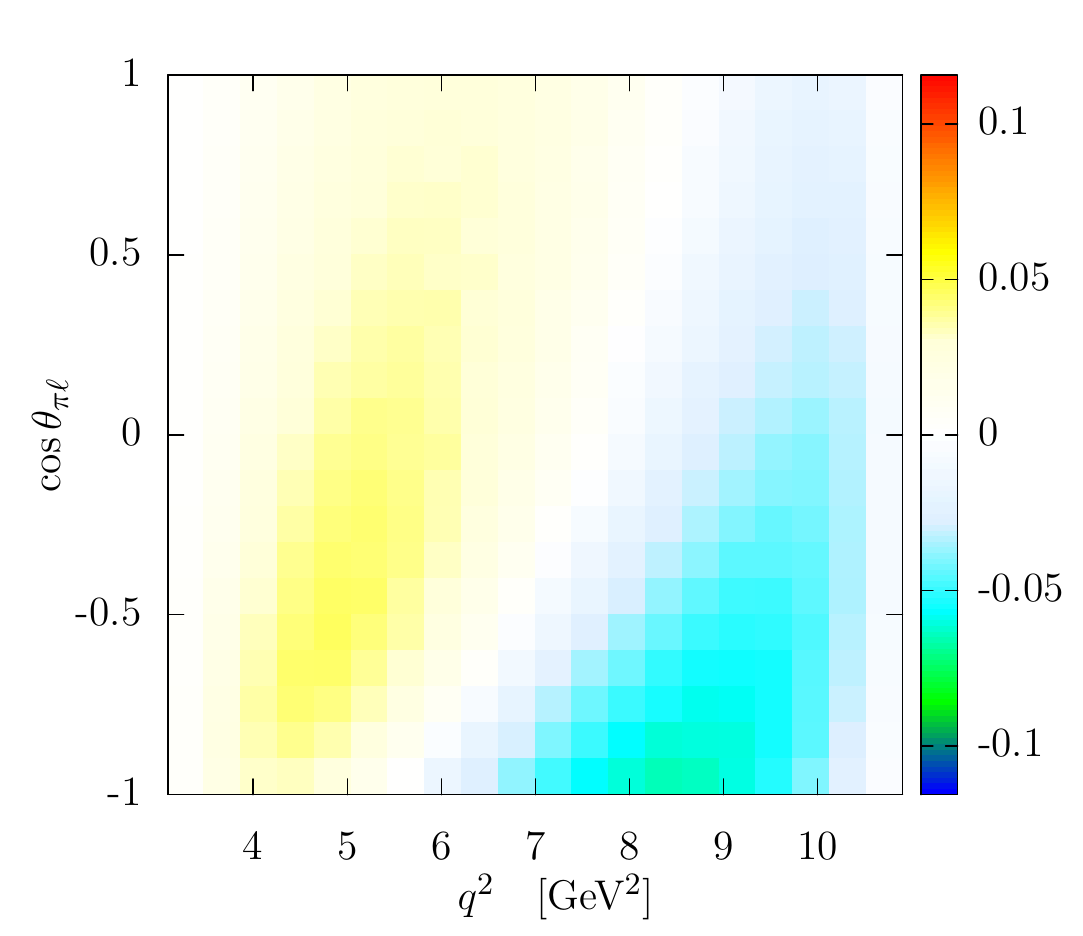}\hfill
\includegraphics[width = 0.32\linewidth]{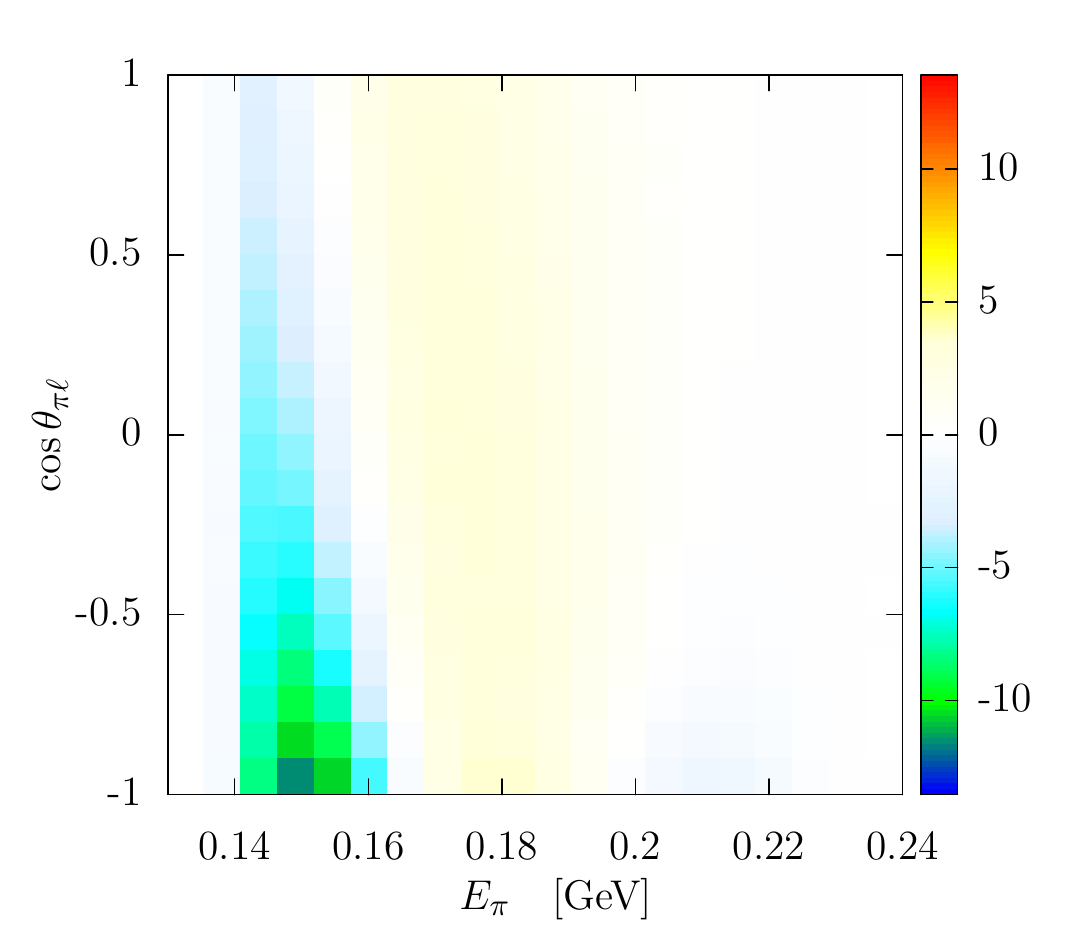} 
\caption{Density contours of $(1/\Gamma) d^2\Gamma/dx\,dy$ for three pairs of kinematic observables, for the SM (top row), $g_T = -0.38$ (middle row) and their difference (bottom row).}
\label{fig:2DH}
\end{figure}

To explore this further, in Fig.~\ref{fig:2DH} we present density plots of the doubly differential decay rates with respect to three pairs of kinematic observables,
\begin{equation}
	\frac{1}{\Gamma}\frac{d^2 \Gamma}{d q^2\, dE_\ell} \,,\qquad \frac{1}{\Gamma}\frac{d^2 \Gamma}{d q^2\, d \cos\theta_{\pi\ell}}\,,\qquad \text{and} \qquad  \frac{1}{\Gamma}\frac{d^2 \Gamma}{d E_\pi\, d \cos\theta_{\pi\ell}}\,,
\end{equation}
for the SM (top row), $g_T = -0.38$ (middle row), and their difference (bottom row). In particular, the density plots for the difference of $d^2 \Gamma/d q^2\, dE_\ell$ and $d^2 \Gamma/d q^2\, d \cos\theta_{\pi\ell}$ have non-trivial level contours, suggesting that an analysis using both of these observables may have significantly more SM--NP discrimination power than $q^2$ or any other single kinematic observable. (A preliminary multivariate study of all ten observables with a boosted decision tree trained to discriminate the SM and the $g_T = -0.38$ model supports this claim~\cite{Hammer:2016}.)

To roughly quantify the relative discrimination power of single and doubly differential distributions in the $q^2$\,--\,$E_\ell$ space, we proceed to divide the MC sample into two bins --- a ``2-binning'' ---  according to a partitioning in each of the one-dimensional $q^2$ and $E_\ell$ distributions as well as in the two-dimensional $q^2$\,--\,$E_\ell$ parameter space. We choose these partitionings at intersection points of contours of the SM and $g_T = -0.38$ theories, to maximize their difference in each bin. From Figs.~\ref{fig:1DH} and \ref{fig:2DH}, this corresponds to 2-binning on either side of
\begin{equation}
	\label{eqn:TBP}
	q^2 \simeq 7.25~\text{GeV}^2\,,\qquad E_\ell \simeq 0.9~\text{GeV}\,,\quad \text{and} \qquad E_\ell \simeq 2.3~\text{GeV} - 0.21~\text{GeV}^{-1}\, q^2 \,.
\end{equation}
The latter partition is shown by a gray dashed line on the $q^2$\,--\,$E_\ell$ difference plot in the bottom left panel in Fig.~\ref{fig:2DH}. 

For each 2-binning, we define a discriminator, 
\begin{equation}
	\chi^2 \equiv \sum_{i,j = 1,2}\big(n^{\text{H}}_i - n^{\text{T}}_i\big)\, \frac{1}{\sigma_{ij}^2}\, \big(n^{\text{H}}_j - n^{\text{T}}_j\big)\,,
\end{equation}
where $n_{1,2}$ are the two bin entries, T (H) labels the true (hypothesis) theory, and $\sigma^2$ is a $2\times2$ covariance matrix.   An approximate covariance matrix for the three 2-binnings is constructed based on the distributions presented in Ref.~\cite{Huschle:2015rga}, measured in a signal-rich region approximated by the phase space cuts~\eqref{eqn:PSC}. We decompose the covariance matrix as
\begin{equation}
	\label{eqn:ACV}
	\sigma^{2} = \sigma_{\text{data}}^{2} + \sigma_{\text{bg}}^{2} + \sigma_{\text{sys}}^{2} + \sigma_{\text{shape}}^{2}\,,
\end{equation}
where we have suppressed the indices. The first term, $\sigma_{\text{data}}^{2}$, corresponds to the Poisson error of the measured data in each bin, while $\sigma_{\text{bg}}^{2}$ corresponds to the error in the normalizations of the main background components, mainly the $D^{**}$ backgrounds, which are fixed by data in different kinematic regions. Both terms therefore scale with the square root of the luminosity. Rescaling statistics to a initial benchmark luminosity of $5$~ab$^{-1}$ at Belle~II implies $\sigma_{\text{data}} \simeq 10\%$ and $\sigma_{\text{bg}} \simeq 14\%$. While $\sigma_{\text{data}}$ is uncorrelated by construction, we assume $\sigma_{\text{bg}}$ is purely an error in overall normalization, and therefore fully correlated between the two bins. By looking at the systematic error breakdown in Ref.~\cite{Huschle:2015rga}, we divide the systematic components into a fully correlated systematic error $\sigma_{\text{sys}} $ and a component $\sigma_{\text{shape}}$ coming from $D^{**}$ background shape variations of unknown correlation between the two bins.
We conservatively assume that systematic errors remain the same in the future, therefore setting $\sigma_{\text{sys}} \sim 4\%$ and $\sigma_{\text{shape}} \sim 3\%$. We emphasize that translation of the $\chi^2$ values, obtained from this approximate covariance matrix~\eqref{eqn:ACV}, into statistical confidence levels requires a more comprehensive treatment of backgrounds and their correlations than attempted here, beyond the scope of the present work. However, the relative size of $\chi^2$ values for different $2$-binnings is less sensitive to background correlation effects, and therefore can be thought of as a proxy for the ratio of the actual $\chi^2$ statistics.

\begin{figure}[t]
	\includegraphics[width= 7cm]{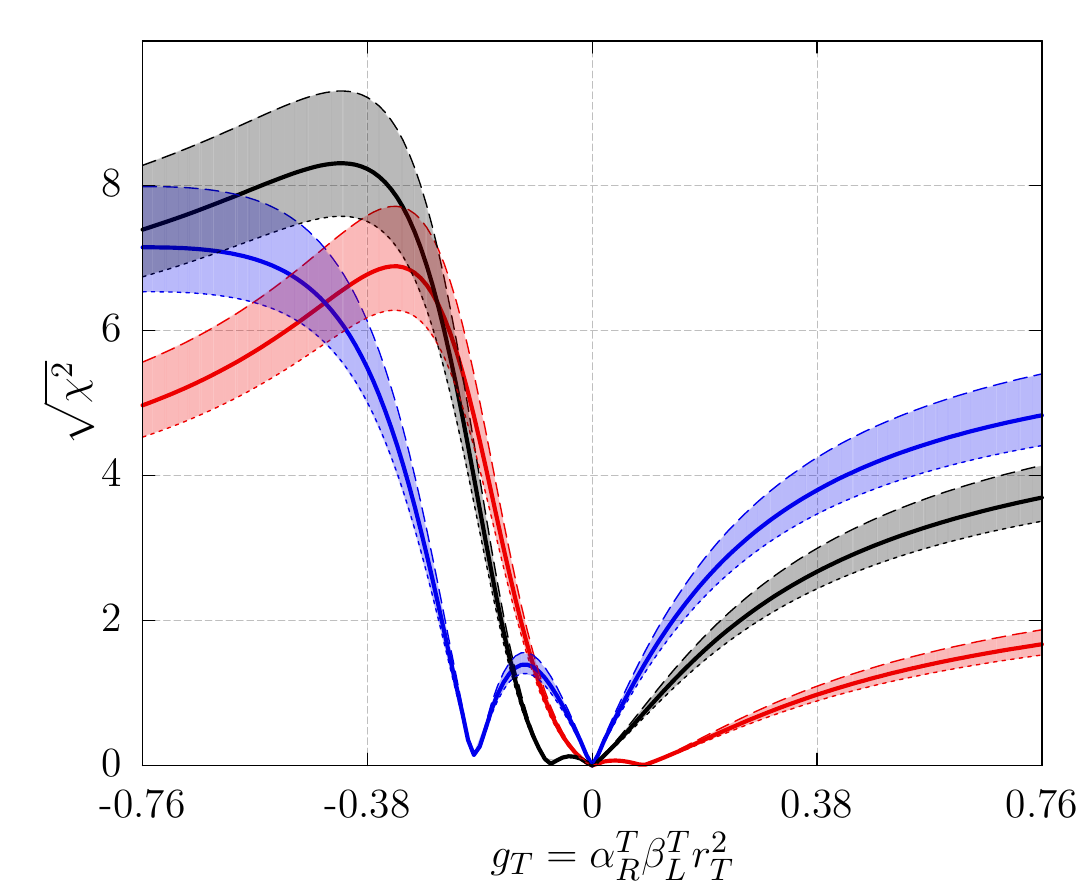} \hfill
	\includegraphics[width= 7cm]{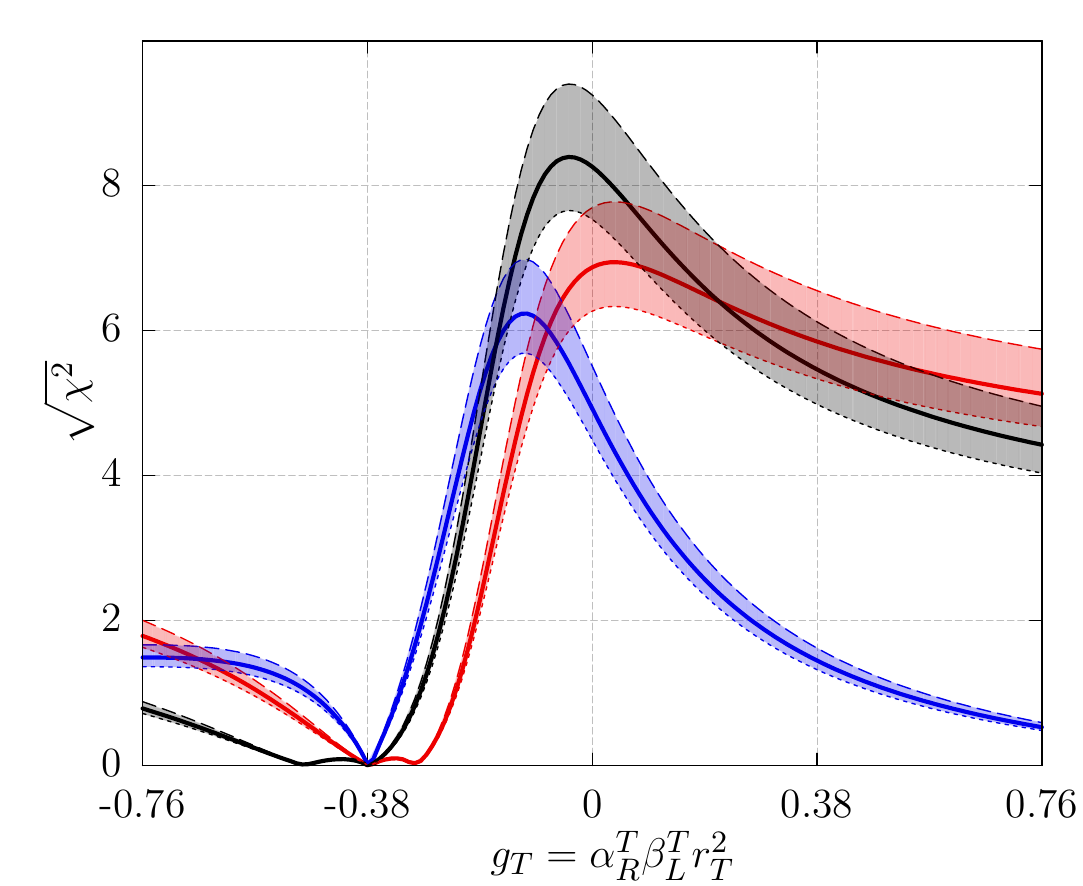} 
	\caption{Approximate $\chi^2$ bands, ranging over arbitrary systematic $\sigma_{\text{shape}}$ (anti)correlations, for 2-binning in $q^2$ (red), $E_\ell$ (blue) and $q^2$--$E_\ell$ (black), according to the partitionings in eq.~\eqref{eqn:TBP}, for the true theory being the SM (left) and $g_T = -0.38$ (right). The phase space cuts in eq.~\eqref{eqn:PSC} are applied, and statistics is rescaled to a future $5$~ab$^{-1}$ luminosity.  Also shown for each 2-binning are contours for uncorrelated (solid), fully correlated (dashed) and fully anticorrelated (dotted) $\sigma_{\text{shape}}$. These $\chi^2$ values are not statistical confidence levels; see text for details.}
	\label{fig:XSS}
\end{figure}

As an example, we now suppose either the SM or the $g_T = -0.38$ model to be the true theory, and consider the space of hypotheses $g_T = [-0.76, 0.76]$. In Fig.~\ref{fig:XSS} we show corresponding $\chi^2$ bands for both theories, generated by ranging over arbitrary correlation for $\sigma_{\text{shape}}$, with phase space cuts~\eqref{eqn:PSC}. We see in Fig.~\ref{fig:XSS} that the two-dimensional $2$-binning for the SM ($g_T = -0.38$) true theory excludes the $g_T = -0.38$ (SM) hypothesis with greater confidence than either of the single observable 2-binnings alone. However, for $g_T$ hypothesis ranges closer to the true theory values, the lepton energy $E_\ell$ 2-binning has greater distinguishing power. An optimized discrimination of these theories using a multivariate analysis will be studied elsewhere.

\section{Summary}
\label{sec:sum}

In this paper we have derived explicit and compact expressions for the $1 \to 4$, $5$ and $6$ body helicity amplitudes for $B \to \dds(\to DY)\tau(\to X\tnbar)\tn$, with $Y = \pi$ or $\gamma$ and $X = \ell \nu_\ell$ or $\pi$, including arbitrary NP contributions from the maximal set of ten four-Fermi operators. These results properly account for interference effects in the full phase space of the $\tau$ and $D^*$ decay products. The former are formally $\mathcal{O}(m_\tau/m_B)$ in the SM, but can be $\mathcal{O}(1)$ in the presence of new physics, and the latter are typically $\mathcal{O}(1)$. While these effects are included in \texttt{EvtGen} for the SM, they are missing from previous NP analyses. This amplitude-level calculation also permits efficient computation of the event weights themselves, which in turn permits efficient reweighting of the large fully simulated MC datasets required for the high statistics analyses at Belle~II and LHCb. 

As an example, we have presented a preliminary exploration of kinematical effects in the phase space of $B \to \dds(\to D\pi)\tau(\to \ell\nu_\ell\tnbar)\tn$ for a class of theories with a NP antisymmetric tensor current. Our amplitude-level calculation makes it feasible to efficiently compute an event `weight matrix' in the space of NP couplings, so that reweighting of the MC dataset need be performed only once per data sample. In this way, not only single but also multidimensional distributions can be rapidly computed for any NP theory. We find that bivariate analyses can exhibit greater discriminating power of the SM versus NP~models. 

Directions for future study include computing the analogous helicity amplitudes for $B \to D^{**}\tau \tn$ using recent form factor results~\cite{Bernlochner:2016bci}, in order to examine the interference effects from the $\tau$ and $D^{**}$  decays. One might also extend the bivariate analysis to consider the hadronic $\tau \to \pi \nu$ mode, given recent results using single kinematic variables~\cite{Abdesselam:2016xqt}. Employment of a boosted decision tree to perform a complete multivariate analysis of the full phase space is also planned. A comprehensive treatment of backgrounds and detector effects will permit estimation of the corresponding statistical confidence levels and future NP exclusion limits achievable with such multivariate analyses at current and upcoming experiments. A software package, \texttt{Hammer}~\cite{Hammer:2016}, is under development, which can be incorporated into existing software pipelines that account for these background and detector effects.

\acknowledgments
We thank Florian Bernlochner and Stephan Duell for helpful conversations and collaboration on \texttt{Hammer}, and Aneesh Manohar for comments on the manuscript. We thank the Aspen Center for Physics, supported by the NSF Grant No.~PHY-1066293, for hospitality while parts of this work were completed. This work was supported in part by the Office of Science, Office of High Energy Physics, of the U.S.\ Department of Energy under contract DE-AC02-05CH11231, and by the National Science Foundation under grant No. PHY-1002399. 
This research used resources of the National Energy Research Scientific Computing Center, which is supported by the Office of Science of the U.S.\ Department of Energy under Contract No.\ DE-AC02-05CH11231. DR acknowledges support from the University of Cincinnati. 

\appendix

\section{Helicity angle expressions}
\label{app:HAE}
In this Appendix we provide expressions for the physical helicity angles in terms of Lorentz invariant combinations of particle momenta. The polar angles $\theta_{D,\tau,W,\ell} \in [0,\pi)$, so we need specify only the cosine of these angles,
\begin{subequations}
\label{eqn:CPHA}
\begin{align}
	\cos\thtau & = \frac{E_W^*}{\Pw \dotpr{k_{\tn}}{q} \dotpr{p_B}{q}} \Big[\dotpr{p_B}{q}\dotpr{k_{\tn}}{q} - q^2\dotpr{p_B}{k_{\tn}} \Big]\,, \\[5pt]
	\cos\thW & = \frac{\dotpr{p_\tau}{k_{\tnbar}} \dotpr{p_\tau}{k_{\tn}} - m_\tau^2 \dotpr{k_{\tn}}{k_{\tnbar}}}{ \dotpr{p_\tau}{k_{\tnbar}} \dotpr{p_\tau}{k_{\tn}} }	\,, \\[5pt]
	\cos\thell & = \frac{2\dotpr{(k_\ell - k_{\nu_\ell})}{k_{\tnbar} } } {m_\tau^2 - p^2}\,,
\end{align}
\end{subequations}
and for $\bdstn$ processes 
\begin{subequations}
\begin{align}
	\cos\thD\Big|_{D^* \to D\pi}  & = \frac{E_{\pi}^*\big(E_{D^*}^*E_W^* + \Pw^2 \big) - m_{D^*}\dotpr{p_\pi} {q} }{ m_B \Pp \Pw} \,,\\[10pt]
	\cos\thD\Big|_{D^* \to D\g}  & = \frac{ \dotpr{k_\g}{p_D}\dotpr{p_{D^*}}{q} -m_{D^*}^2\dotpr{k_\g}{q} }{ m_B \Pw\dotpr{k_\g}{p_D} }\,.
\end{align}
\end{subequations}
Note that $\cos\thW$ is defined with $p$ dependence implicit, so that for $\tau \to \pi \tnbar$ one need only replace $\thW \to \thpi$ in eq.~\eqref{eqn:CPHA}. In these expressions, the $B$ rest frame energies
\begin{equation}
	E_{W}^* = \frac{m_{B}^2 - m_{D^*}^2 + q^2}{2m_{B}}\,,\qquad E_{D^*}^* =  \frac{m_{B}^2  - q^2 + m_{D^*}^2}{2m_{B}}\,,
\end{equation}
and the $D^*$ rest frame energy $E_{\pi}^* = (m_{D^*}^2 - m_{D}^2 + m_{\pi}^2)/2m_{D^*}$.

For the azimuthal angles, only the combinations $\phtau-\phW$, $\phW - \phell$ and $\phD - \phtau$ appear in the helicity amplitudes. We therefore provide direct expressions for the sine and cosine of these relative twist angles, rather than for the azimuthal helicity angles themselves. To keep expressions short, we express these twist angles iteratively in terms of trigonometric functions of the polar helicity angles,
\begin{subequations}
	\begin{align}
		\sin(\phtau - \phW) 
			& = -\frac{\sqrt{q^2}\tan^2[\thW/2]\,\epspr{p_B}{p_{\dds}}{k_{\tn}}{k_{\tnbar}} }{ m_B m_\tau \Pw \sin \thtau \dotpr{k_{\tn}}{k_{\tnbar}}}\,,\\[5pt]
		\cos(\phtau - \phW)
			 & = \frac{\sqrt{q^2} \csc \thtau \csc \thW }{m_B m_\tau \Pw \dotpr{p_\tau}{k_{\tnbar}}\dotpr{p_\tau}{k_{\tn}}} 
			 		\Big\{ \dotpr{p_\tau}{k_{\tn}}\big[m_\tau^2 \dotpr{p_B}{k_{\tnbar}}  - \dotpr{p_B}{p_\tau}\dotpr{p_\tau}{k_{\tnbar}} \big] \notag\\
			 & \qquad - \cos\thW  \dotpr{p_\tau}{k_{\tnbar}} \big[m_\tau^2 \dotpr{p_B}{k_{\tn}}  - \dotpr{p_B}{p_\tau}\dotpr{p_\tau}{k_{\tn}} \big]\Big\}\,,\\[5pt]
		\sin(\phell - \phW) 
			& =  \frac{2\tan[\thW/2]\,\epspr{k_\ell}{k_{\nu_\ell}}{k_{\tn}}{k_{\tnbar}}}{m_\tau \sqrt{p^2} \sin \thell \dotpr{k_{\tn}}{k_{\tnbar}}}\,,\\[5pt]
		\cos(\phell-\phW) 
			& = \frac{\csc \thell \csc \thW }{m_\tau \sqrt{p^2}\dotpr{p_\tau}{k_{\tn}}}
					\Big\{ m_\tau^2\big[ 2 \dotpr{k_{\nu_\ell}}{k_{\tn}} + (\cos\thell\cos\thW -1)\dotpr{p_\tau}{k_{\tn}} \big] \notag \\
			& \qquad + (1 - \cos\thW)(1+ \cos\thell) \dotpr{p_\tau}{k_{\tn}} \dotpr{p_\tau}{k_{\tnbar}}\Big\}\,,		
	\end{align}
\end{subequations}
with $\epsilon^{0123} = +1$, and for $\bdstn$ processes 
\begin{subequations}
\begin{align}
		\sin(\phD - \phtau)\Big|_{D^* \to D\pi} & = \frac{\sqrt{q^2}\csc \thD \csc \thtau \epspr{p_B}{p_D}{p_\pi}{k_{\tn}}}{m_B \Pp \Pw \dotpr{p_\tau}{k_{\tn}}}\,, \\[10pt]
		\sin(\phD - \phtau)\Big|_{D^* \to D\g} & = \frac{m_{D^*}\sqrt{q^2} \csc \thD \csc \thtau \epspr{p_B}{p_D}{k_{\g}}{k_{\tn}}}{m_B \Pw \dotpr{k_\g}{p_D} \dotpr{p_\tau}{k_{\tn}}} \,,\\[10pt]
		\cos(\phD - \phtau)\Big|_{D^* \to D\pi} & = -\frac{\csc \thD \csc \thtau}{m_{D^*}\Pp \sqrt{q^2} \dotpr{k_{\tn}}{q}}\Big\{E_\pi^*\big[q^2 \dotpr{p_B}{k_{\tn}} - \dotpr{q}{k_{\tn}} \dotpr{p_B}{q}\big] \\
					&\qquad + m_{D^*}\big[\dotpr{q}{k_{\tn}} \dotpr{p_\pi}{q} - q^2 \dotpr{p_\pi}{k_{\tn}} \big] + \Pp \cos \thD \cos \thtau  \dotpr{p_{D^*}}{q} \dotpr{k_{\tn}}{q}\Big\}\,, \notag \\[10pt]
		\cos(\phD - \phtau)\Big|_{D^* \to D\g} & = \frac{\csc \thD \csc \thtau}{m_{D^*}\sqrt{q^2}\dotpr{q}{k_{\tn} } \dotpr{p_D}{k_\g} }\Big\{m_{D^*}^2\big[q^2 \dotpr{k_\g}{k_{\tn}} -\dotpr{q}{k_{\g}}\dotpr{q}{k_{\tn}} \big] \notag \\
					 & \qquad + \dotpr{k_\g}{p_D}\dotpr{q}{k_{\tn}} \Big[\dotpr{p_B}{q}(1 + \cos\thD \cos \thtau) \notag \\
					 & \qquad -(m_B^2 -m_{D^*}^2) \cos \thD \cos \thtau \Big] - q^2 \dotpr{p_B}{k_{\tn}}\dotpr{k_\g}{p_D}\Big\} \,.
\end{align}
\end{subequations}

\section{\texorpdfstring{$B \to D^{*}(\to D \g)\tau \tn$}{bdsdgtn}}
\label{app:DDG}
For $B \to D^{*} (\to D \g)\tau \tn$, the helicity amplitudes $[\mathcal{A}_{B \to D^{*}(\to D \g)\tau \tn}]^{\kappa s_{\tn}}_{s_{\tau}} \equiv [\mathcal{A}^\g]^{\kappa s_{\tn}}_{s_\tau}$ obey a parity relation
\begin{equation}
	\label{eqn:DGPR}
	[\mathcal{A}^\g]^{\pm s_{\tn}}_{s_\tau}(\thD) = [\mathcal{A}^\g]^{\mp s_{\tn}}_{s_\tau}(\thD + \pi)\,.
\end{equation}
Hence, one need only explicitly express half of the helicity amplitudes. 

The decay $D^* \to D\g$ proceeds via the operator $(e \mu_a/4)\epsilon^{\mu\nu\rho\sigma} (\partial_\mu D^*_{\nu}-\partial_\nu D^*_{\mu} ) F_{\rho\sigma} D$, in which, following the notation of Ref.~\cite{Manohar:2000dt}, $\mu_a$ is a magnetic moment such that
\begin{equation}
	e \mu_a =  \bigg[12 \pi \Gamma(D^* \to D\g) \frac{ 8 m_{D^*}^3}{(m_{D^*}^2 - m_D^2)^3}\bigg]^{1/2}\,.
\end{equation}
We define the functions
\begin{subequations}
\begin{align}
	\DgaA_\pm & \equiv \sin^2\frac{\thD}{2} \cos^2\frac{\thtau}{2}e^{-i \phDtau} \pm \cos^2\frac{\thD}{2} \sin^2\frac{\thtau}{2}e^{ i \phDtau} \,, \\
	\DgaA_0 & \equiv \sin\thD \sin \thtau\,,\\
	\DgaB_\pm & \equiv \sin\thtau\bigg[ \cos^2\frac{\thD}{2}e^{i\phDtau} \pm \sin^2\frac{\thD}{2}e^{-i\phDtau} \bigg]\,,\\
	\DgaB_0 & \equiv \sin\thD\cos\thtau\,,\\
	\DgaB_D & \equiv \sin\thD\,.
\end{align}
\end{subequations}
The $\DgaA$ and $\DgaB$ functions play the same role as $\DpiA$ and $\DpiB$ in the $D^* \to D\pi$ mode above. That is, the $s_\tau = 2$ ($s_\tau = 1$) helicity amplitudes are linear combinations of the $\DgaA$ ($\DgaB$) functions exclusively. Each set of $\DgaA$ and $\DgaB$ functions is $L^2(\mathbb{C})$ orthogonal under integration over the angular phase space $d\Omega_D d \Omega_\tau $, while the $\DgaA$ functions are orthogonal with respect to $\DpiB$ with the inclusion of an additional $e^{\pm i \phtau}$ phase in the integration measure, in accordance with our $\tau$ spinor phase conventions~\eqref{eqn:TSBC}. One finds
\begin{subequations}
\begin{align}
& [\mathcal{A}^\g]^{+\dn}_1  =   -2i V_{cb} e \mu_a G_F (m_{D^*}^2 - m_D^2) \sqrt{q^2 - m_{\tau}^2}\bigg\{ \\ 
    & +\frac{i a_0(q^2)  m_{B} \Pw  (-\alSL +\alSR ) \beSL  r_{S}^2 \DgaB_D }{4 m_{D^*}} \notag \\ 
    & +i f(q^2)  m_{\tau} (-1 + (\alVR - \alVL)\beVL r_V^2) \bigg[\frac{(- m_B^2 + m_{D^*}^2 + q^2) \DgaB_0 }{8 m_{D^*} q^2 }+\frac{m_{B} \Pw  \DgaB_D }{4 m_{D^*} q^2 }+\frac{\DgaB_- }{4 \sqrt{q^2 }}\bigg] \notag \\ 
    & +\frac{i g(q^2)  m_{B} m_{\tau} \Pw  (1 + (\alVL + \alVR)\beVL r_V^2) \DgaB_+ }{2 \sqrt{q^2 }} -\frac{i a_-(q^2)  m_{B} m_{\tau} \Pw  (1 + (\alVL - \alVR)\beVL r_V^2) \DgaB_D }{4 m_{D^*}} \notag \\ 
    & +i a_+(q^2)  m_{B} m_{\tau} \Pw  (1 + (\alVL - \alVR)\beVL r_V^2) \bigg[\frac{m_{B} \Pw  \DgaB_0 }{2 m_{D^*} q^2 }+\frac{(- m_B^2 + m_{D^*}^2) \DgaB_D }{4 m_{D^*} q^2 }\bigg] \notag \\ 
    & -\frac{2 i a_{T_0}(q^2)  m_{B}^2 \Pw ^2 \alTR  \beTL  r_{T}^2 \DgaB_0 }{m_{D^*}} +i a_{T_-}(q^2)  \alTR  \beTL  r_{T}^2 \bigg[\frac{(- m_B^2 + m_{D^*}^2 + q^2) \DgaB_0 }{2 m_{D^*}}+\sqrt{q^2 } \DgaB_- \bigg] \notag \\ 
    & +i a_{T_+}(q^2)  \alTR  \beTL  r_{T}^2 \bigg[\frac{2 m_{B} \Pw  \DgaB_+ }{\sqrt{q^2 }}-\frac{(m_B^2 + 3m_{D^*}^2 - q^2) \DgaB_0 }{2 m_{D^*}}+\frac{(m_B^2 - m_{D^*}^2) \DgaB_- }{\sqrt{q^2 }}\bigg]\bigg\}\notag\\[15pt]
& [\mathcal{A}^\g]^{+\dn}_2  =  -2i V_{cb} e \mu_a G_F (m_{D^*}^2 - m_D^2) \sqrt{q^2 - m_{\tau}^2}\bigg\{ \\ 
    & -i f(q^2)  (-1 + (\alVR - \alVL)\beVL r_V^2) \bigg[\frac{\DgaA_+ }{2}+\frac{(- m_B^2 + m_{D^*}^2 + q^2) \DgaA_0 }{8 m_{D^*} \sqrt{q^2 }}\bigg] \notag \\ 
    & +i g(q^2)  m_{B} \Pw  (1 + (\alVL + \alVR)\beVL r_V^2) \DgaA_-   -\frac{i a_+(q^2)  m_{B}^2 \Pw ^2 (1 + (\alVL - \alVR)\beVL r_V^2) \DgaA_0 }{2 m_{D^*} \sqrt{q^2 }} \notag \\ 
    & +\frac{2 i a_{T_0}(q^2)  m_{B}^2 m_{\tau} \Pw ^2 \alTR  \beTL  r_{T}^2 \DgaA_0 }{m_{D^*} \sqrt{q^2 }}  -i a_{T_-}(q^2)  m_{\tau} \alTR  \beTL  r_{T}^2 \bigg[2 \DgaA_+ +\frac{(- m_B^2 + m_{D^*}^2 + q^2) \DgaA_0 }{2 m_{D^*} \sqrt{q^2 }}\bigg] \notag \\ 
    & +i a_{T_+}(q^2)  m_{\tau} \alTR  \beTL  r_{T}^2 \bigg[-\frac{2 (m_B^2 - m_{D^*}^2) \DgaA_+ }{q^2 }+\frac{(m_B^2 + 3m_{D^*}^2 - q^2) \DgaA_0 }{2 m_{D^*} \sqrt{q^2 }}+\frac{4 m_{B} \Pw  \DgaA_- }{q^2 }\bigg]\bigg\}\notag\\[15pt]
& [\mathcal{A}^\g]^{-\up}_1  = -2i V_{cb} e \mu_a G_F (m_{D^*}^2 - m_D^2) \sqrt{q^2 - m_{\tau}^2}\bigg\{ \\ 
    & +i f(q^2)  (-\alVL +\alVR ) \beVR  r_{V}^2 \bigg[\frac{(- m_B^2 + m_{D^*}^2 + q^2) \DgaA_0 }{8 m_{D^*} \sqrt{q^2 }}+\frac{\DgaA_+^* }{2}\bigg] \notag \\ 
    & +i g(q^2)  m_{B} \Pw  (\alVL +\alVR ) \beVR  r_{V}^2 \DgaA_-^*  +\frac{i a_+(q^2)  m_{B}^2 \Pw ^2 (\alVL -\alVR ) \beVR  r_{V}^2 \DgaA_0 }{2 m_{D^*} \sqrt{q^2 }} \notag \\ 
    & +\frac{2 i a_{T_0}(q^2)  m_{B}^2 m_{\tau} \Pw ^2 \alTL  \beTR  r_{T}^2 \DgaA_0 }{m_{D^*} \sqrt{q^2 }} -i a_{T_-}(q^2)  m_{\tau} \alTL  \beTR  r_{T}^2 \bigg[\frac{(- m_B^2 + m_{D^*}^2 + q^2) \DgaA_0 }{2 m_{D^*} \sqrt{q^2 }}+2 \DgaA_+^* \bigg] \notag \\ 
    & +i a_{T_+}(q^2)  m_{\tau} \alTL  \beTR  r_{T}^2 \bigg[\frac{(m_B^2 + 3m_{D^*}^2 - q^2) \DgaA_0 }{2 m_{D^*} \sqrt{q^2 }}-\frac{2 (m_B^2 - m_{D^*}^2) \DgaA_+^* }{q^2 }+\frac{4 m_{B} \Pw  \DgaA_-^* }{q^2 }\bigg]\bigg\}\notag\\[15pt]
& [\mathcal{A}^\g]^{-\up}_2  =  -2i V_{cb} e \mu_a G_F (m_{D^*}^2 - m_D^2) \sqrt{q^2 - m_{\tau}^2}\bigg\{ \\ 
    & +\frac{i a_0(q^2)  m_{B} \Pw  (-\alSL +\alSR ) \beSR  r_{S}^2 \DgaB_D }{4 m_{D^*}} \notag \\ 
    & +i f(q^2)  m_{\tau} (-\alVL +\alVR ) \beVR  r_{V}^2 \bigg[\frac{(- m_B^2 + m_{D^*}^2 + q^2) \DgaB_0 }{8 m_{D^*} q^2 }+\frac{m_{B} \Pw  \DgaB_D }{4 m_{D^*} q^2 }+\frac{\DgaB_-^* }{4 \sqrt{q^2 }}\bigg] \notag \\ 
    & -\frac{i g(q^2)  m_{B} m_{\tau} \Pw  (\alVL +\alVR ) \beVR  r_{V}^2 \DgaB_+^* }{2 \sqrt{q^2 }}  +\frac{i a_-(q^2)  m_{B} m_{\tau} \Pw  (-\alVL +\alVR ) \beVR  r_{V}^2 \DgaB_D }{4 m_{D^*}} \notag \\ 
    & +i a_+(q^2)  m_{B} m_{\tau} \Pw  (\alVL -\alVR ) \beVR  r_{V}^2 \bigg[\frac{m_{B} \Pw  \DgaB_0 }{2 m_{D^*} q^2 }+\frac{(- m_B^2 + m_{D^*}^2) \DgaB_D }{4 m_{D^*} q^2 }\bigg] \notag \\ 
    & +\frac{2 i a_{T_0}(q^2)  m_{B}^2 \Pw ^2 \alTL  \beTR  r_{T}^2 \DgaB_0 }{m_{D^*}}  -i a_{T_-}(q^2)  \alTL  \beTR  r_{T}^2 \bigg[\frac{(- m_B^2 + m_{D^*}^2 + q^2) \DgaB_0 }{2 m_{D^*}}+\sqrt{q^2 } \DgaB_-^* \bigg] \notag \\ 
    & +i a_{T_+}(q^2)  \alTL  \beTR  r_{T}^2 \bigg[\frac{(m_B^2 + 3m_{D^*}^2 - q^2) \DgaB_0 }{2 m_{D^*}}-\frac{2 m_{B} \Pw  \DgaB_+^* }{\sqrt{q^2 }}+\frac{(- m_B^2 + m_{D^*}^2) \DgaB_-^* }{\sqrt{q^2 }}\bigg]\bigg\}\,, \notag
\end{align}
\end{subequations}
with $r_{V,S,T} \equiv m_W/\Lambda_{V,S,T}$. The four remaining helicity amplitudes $[\mathcal{A}^\g]^{-\dn}_{s_\tau}$ and $[\mathcal{A}^\g]^{+\up}_{s_\tau}$ follow immediately from the parity relation~\eqref{eqn:DGPR}.

%

\end{document}